\documentclass[prb,twocolumn,superscriptaddress,showpacs]{revtex4}
\usepackage{graphicx}% Include figure files
\usepackage{dcolumn}% Align table columns on decimal point
\usepackage{bm}% bold math
\usepackage{xcolor}
\usepackage{soul}
\usepackage{graphicx}% Include figure files
\usepackage{dcolumn}% Align table columns on decimal point
\usepackage{bm}% bold math
\usepackage{amsmath}
\begin{document}

\title{Coexistence of giant Cooper pairs with a bosonic condensate and anomalous behavior of energy gaps in the BCS-BEC crossover of a two-band superfluid Fermi gas}
\author{Yuriy Yerin}
\affiliation{School of Science and Technology, Physics Division, Universit\`{a} di Camerino, 62032 Camerino (MC), Italy}
\author{Hiroyuki Tajima}
\affiliation{Quantum Hadron Physics Laboratory, RIKEN Nishina Center, Wako, Saitama, 351-0198, Japan}
\author{Pierbiagio Pieri}
\affiliation{School of Science and Technology, Physics Division, Universit\`{a} di Camerino, 62032 Camerino (MC), Italy}
\affiliation{INFN, Sezione di Perugia, 06123 Perugia (PG), Italy}
\author{Andrea Perali}
\affiliation{School of Pharmacy, Physics Unit, Universit\`{a} di Camerino, 62032 Camerino (MC), Italy}

\begin{abstract}
We investigate Bardeen-Cooper-Schrieffer (BCS) $-$ Bose-Einstein condensation (BEC) crossover in a two-band superfluid Fermi gas with an energy shift between the bands. When the intraband coupling in the cold (first) band is fixed as weak, we find that in the case of vanishing interband interaction and in the strong-coupling limit of the hot (second) band the system undergoes a transition to a single-component configuration with the full suppression of the first energy gap and with the full redistribution of particles between bands. For non-vanishing interband interaction we reveal the non-monotonic dependence of the energy gap in the first band vs intraband coupling in the second band with the presence of a  hump. 
In the case of weak interband coupling the system shows a significant amplification of the intrapair correlation length of the condensate in the first band in the strong-coupling regime of the second band, which clearly indicates the coexistence of giant Cooper pairs and a bosonic condensate even for nonzero temperatures. This can lead to a non-monotonic temperature dependence of the second energy gap with a peak. 
Here predicted coexistence of the giant Cooper pairs and bosonic molecules can be verified by means of the visualization of vortex cores in the two-component atomic condensates as well as in some iron-based superconductors.
\end{abstract}

\date{\today}

\maketitle

\section{Introduction}

A model of two-band superfluidity has been considered for a long time solely as the next iteration step to the BCS theory of superconducting state to take into account the anisotropic properties of metals and the effect of overlapping of the energy bands in the vicinity of their Fermi surface, which leads to the appearance of interband quantum electron transitions and, as a result, to an additional indirect interaction between the electrons of each band \cite{Suhl, Moskalenko}. The explosive growth in the study of multiband superconductivity began from the discovery of unconventional superconductivity with a complex structure of the superconducting order parameter (cuprates, heavy-fermion compounds, borocarbides, fullerides, strontium ruthenate, organic superconductors,  iron pnictides and chalcogenides). Complex structure of the order parameter gives rise  to a much richer nomenclature of topological objects and effects in unconventional superconductors in comparison with their conventional counterparts.  These superconducting systems can lead to the formation of a variety of quantum phenomena: states that broke time-reversal symmetry (BTRS), new collective modes, phase domains, vortices with fractional flux, and fractional Josephson effect \cite{S.-Z. Lin, Tanaka, Milosevic, Omelyanchouk}, and shape resonance in the superconducting properties \cite{Bianconi1, Bianconi2}.

Another intriguing aspect is the fact that compounds with unconventional superconductivity can demonstrate anomalous normal-state properties above their critical temperature, which are interpreted as the pseudogap state. The existence of a pseudogap state has been firstly argued in the context of the crossover from BCS superconductivity to the Bose-Einstein condensation in the ground state and at the finite temperature \cite{NSR, Randeria} for underdoped high-$T_c$ cuprate superconductors  \cite{Perali_2002, Palestini_2012, Marsiglio_2015}. In these compounds, pseudogap formation and non-Fermi liquid behavior are well established, and unusual superconducting fluctuations have also been detected above the critical temperature. However, the pseudogap state appears at a much higher temperature than the onset temperature of superconducting fluctuations. At this moment it is still debatable question whether the system is deep inside the crossover regime and to what extent the crossover physics can be relevant to the phase diagram of underdoped cuprate superconductors. 

A magnesium diboride superconductor \cite{Bianconi3, Bianconi4} and recently discovered family of iron-based superconductors with the multiband electron structure and multiple energy gaps offer a new platform for the experimental observation of the  BCS-BEC crossover, providing an opportunity to study new problems about crossover, fluctuation phenomena and pseudogap  in multi-component systems, which go beyond the single-band physics \cite{Guidini}. For instance, ${\rm{BaF}}{{\rm{e}}_{\rm{2}}}{\left( {{\rm{A}}{{\rm{s}}_{{\rm{1 - x}}}}{{\rm{P}}_{\rm{x}}}} \right)_{\rm{2}}}$ may approach the BCS-BEC crossover regime near a quantum critical point \cite{Hashimoto, Shibauchi}. Another candidate is iron chalcogenide ${\rm{F}}{{\rm{e}}_{{\rm{1 + y}}}}{\rm{S}}{{\rm{e}}_{\rm{x}}}{\rm{T}}{{\rm{e}}_{{\rm{1 - x}}}}$, in which the Fermi energy of FeSe is extremely small and can be tuned by chemically doping through the BCS-BEC crossover \cite{Lubashevsky, Okazaki, Kasahara, Kasahara1} . It was found experimentally that the dimensionless measure of the pairing strength, i.e. the ratio of the energy gap and the Fermi energy  $\Delta/E_F$ = 0.16, 0.3 and 0.5, increases monotonically with decreasing of the iron excess $y$, exhibiting a crossover from the BCS to the BEC regime \cite{Rinott}. The investigation of the vortex core by means of scanning tunneling microscopy (STM) shows the presence of Friedel-like oscillations, confirming the BCS-BEC crossover nature of FeSe and a peculiar missing of the pseudogap \cite{Hanaguri}.

Despite that, for the most of multiband superconducting systems the tuning  of interband or intraband interactions are rather challenging and their properties can not be studied easily away from the BCS regime. 
Strongly interacting superfluid systems can be replicated experimentally with ultracold atomic Fermi gases in optical lattices or in single traps confining clouds of fermionic atoms with several hyperfine states \cite{Köhl, Ospelkaus, Chin}. In such systems the interaction strength is adjusted by means of Fano-Feshbach resonances which allow the evolution of superfluidity throughout the BCS-BEC crossover. The newly realized orbital Feshbach resonance in a $^{{\rm{173}}}{\rm{Yb}}$ Fermi gas promises a new wave for studying two-band Fermi system with Josephson-like interaction between bands, enabling the tuning of inter-orbital interactions based on the Zeeman shift of different nuclear spin states of the atoms \cite{Pagano, Höfer, Zhang}. The many-body Hamiltonian governing the physical properties of alkaline-earth Fermi gases across an orbital Feshbach resonance is similar to that of two-band s-wave superconductors, and the description of the BCS-BEC crossover in these systems requires two components of the order parameter, in contrast to a Fermi gas with a single orbital near the broad magnetic Feshbach resonance. Thus, experimental activity in this direction raises fundamentally new problems about the BCS-BEC crossover in multiband superfluids and calls for the theoretical predictions of possible unusual effects \cite{Iskin1, Iskin2, Iskin3, Iskin4, Reyes1, Mondal, Tajima, Chubukov, Wolf, Salasnich}. At this moment the evolution of low energy collective excitations from BCS to BEC coupling regime in two-band s-wave superfluids coupled via an interband Josephson interaction at $T=0$ has been studied \cite{Iskin1}. Later within a mean-field theory generalized to the case of two bands,  the characteristics of two-band superfluidity throughout the BCS-BEC crossover were analyzed and results have been reported only for coincident bands \cite{Iskin2}.  Furthermore, based on the extension of the Nozi\`{e}res-Schmitt-Rink approach \cite{NSR} for two bands, strong enhancement of the critical temperature, a significant reduction of the preformed pair region where pseudogap effects are expected, and the entanglement of two kinds of composite bosons in the strong-coupling BEC regime were predicted for a two-band attractive Fermi system in the normal state with a shallow band coupled to a weakly-interacting deeper band \cite{Tajima}. 

In this paper, using a mean-field theory for a two-band superfluid with gap equations coupled to the density equation we show that a two-band superfluid Fermi gas with energy shift between the bands reveals unique features of the BCS-BEC crossover, which are not realized in the single-band system. The paper is organized as follows.  In Sec. II, we present the model and the main equations of a mean-field approach for the description of the BCS-BEC crossover in a two-band system. In Sec. III, we provide the results of our numerical calculations for the energy gaps, chemical potential, particle densities and the intrapair correlation lengths and discuss unique features of the BCS-BEC crossover, in particular, a coexistence of giant Cooper pairs and bosonic condensate in the strong-coupling regime. We summarize our conclusions in Sec. IV.  Two appendices with analytical calculations and technical details are reported at the end of the paper. 

\section{Model and basic equations}
We consider a two-band system of interacting fermions in three dimensions (3D), where the two fermionic bands have a parabolic dispersion law
\begin{equation}
\label{eq1}
{\xi _i}\left( {\bf{k}} \right) = \frac{{{{ |\bf{k}} |^2}}}{{2m}} - \mu  + \epsilon_i,
\end{equation}
where $\bf{k}$ is the wave-vector, $m$ the effective mass which is assumed equal for both bands, $\mu$ the chemical potential and  $\epsilon_i$ the energy of the bottom of the bands. The index $i$ = 1, 2 numerates the bands, where $i$ = 1 denotes the lower band and $i$ = 2 is the upper band. We set $\epsilon_1 = 0$ and $\epsilon_2 = E_g$ where the value $E_g$ defines the energy shift between the  two bands of the system (Fig. 1).

\begin{figure}
\includegraphics[width=1.05\columnwidth]{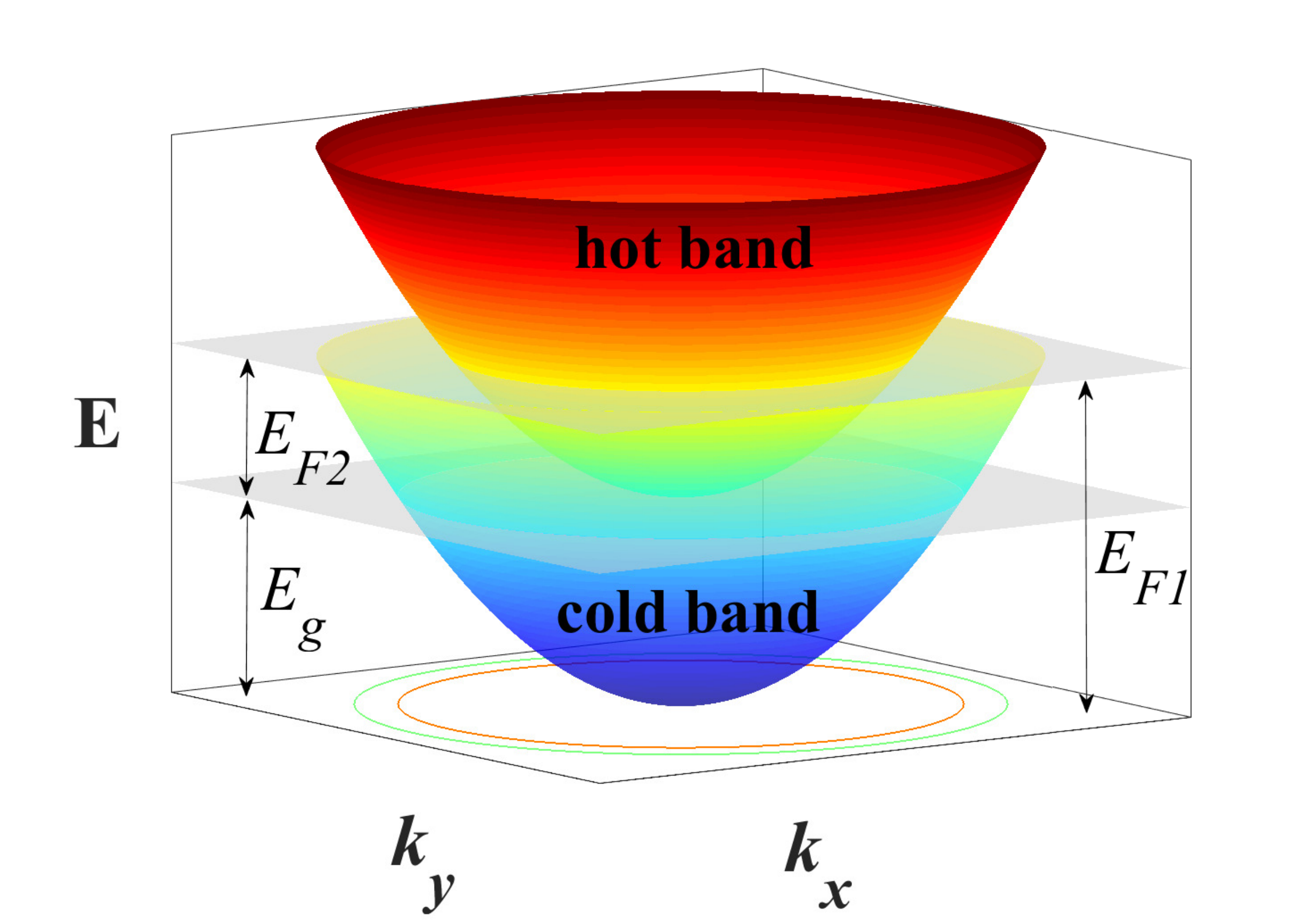}
\caption{The band structure of the two-band superfluid Fermi gas under consideration  ($k_z = 0$ projection). $E_g$ is the energy shift between the 1st ($i=1$) and the 2nd ($i=2$) band. $E_{{\rm F}i}$ corresponds to the Fermi energy of $i$-band in the absence of interactions.}
\label{fig1}
\end{figure}

The effective pairing interaction between fermions is approximated by a separable potential 
\begin{equation}
\label{eq2}
{V_{ij}}\left( {{\bf{k}},{\bf{k'}}} \right) = -{U_{ij}}\Theta \left( {k_0 - |\bf{k}|} \right)\Theta \left( {k_0 -  |\bf{k'} |} \right),
\end{equation}
where $U_{ij}$ are the strength of intraband (when $i=j$) and interband (when $i\neq j$) interactions, $k_0$ is the cut-off momentum, which is supposed the same for intraband and
interband pairing terms and $\Theta \left( x \right)$ is the Heaviside function.  The sign of $U_{12}$ determines the symmetry of the order parameter in the clean case. A repulsive interband interaction constant $U_{12} < 0$ leads to a ground state with $\pi$-phase difference between the two bands, while attractive interband interactions $U_{12} > 0$ stabilize a ground state with a zero-phase difference between their gap functions \cite{Yerin_2007}.

The ground state of the two-band system is examined within a mean-field theory. We generalize single-band approach for a two-band case and write the equations for the energy gaps 
\begin{equation}
\label{eq3}
{\Delta _i}\left( {\bf{k}} \right) =  - \frac{1}{\Omega }\sum\limits_j {\sum\limits_{k'} {{V_{ij}}\left( {{\bf{k}},{\bf{k'}}} \right)} \frac{{{\Delta _j}\left( {{\bf{k'}}} \right)\tanh \frac{{{E_i}\left( {{\bf{k'}}} \right)}}{{2T}}}}{{2{E_i}\left( {{\bf{k'}}} \right)}}}.
\end{equation}
Here $\Omega$ is the volume occupied by the system under consideration, ${E_i}\left( {{\bf{k'}}} \right) = \sqrt {\xi _j^2\left( {{\bf{k'}}} \right) + \Delta _j^2\left( {{\bf{k'}}} \right)} $ are excitation branches in the superfluid state and the gaps having the same cut-off generated by the separable interaction
\begin{equation}
\label{eq4}
{\Delta _i}\left( {\bf{k}} \right) = {\Delta _i}\Theta \left( {k_0 -  |\bf{k}|} \right).
\end{equation}
The coupled equations for the energy gaps must be supplemented with the equation for the total particle density of the system, as the renormalization of the chemical potential is a key feature of the BCS-BEC crossover. 
We consider the total density of particles of the two-band system in the form of an additive contribution from each band 
\begin{equation}
\label{eq5-5}
n = {n_1} + {n_2},
\end{equation}
where $n_i$ is the particle density in each band
\begin{equation}
\label{eq5}
n_i = \frac{2}{\Omega } {\sum\limits_{\bf{k}} {\left[ {v_i^2\left( {\bf{k}} \right)f\left( { - {E_i\left( {\bf{k}} \right)}} \right) + u_i^2\left( {\bf{k}} \right)f\left( {{E_i\left( {\bf{k}} \right)}} \right)} \right]} },
\end{equation}
where $f(z)$ is the Fermi-Dirac distribution function. Here we introduce weights of occupied states via the functions $v_i(\bf{k})$ and $u_i(\bf{k})$ 
\begin{equation}
\label{eq_weight1}
v_i^2\left( {\bf{k}} \right) = \frac{1}{2}\left[ {1 - \frac{{{\xi _i}\left( {\bf{k}} \right)}}{{E_i\left( {\bf{k}} \right) }}} \right],
\end{equation}
\begin{equation}
\label{eq_weight2}
u_i^2\left( {\bf{k}} \right) = 1 - v_i^2\left( {\bf{k}} \right).
\end{equation}
During the calculations $n$ will be taken as a value $n = n_1^0 + n_2^0 = \frac{{k_{F1}^3}}{{3{\pi ^2}}} + \frac{{k_{F2}^3}}{{3{\pi ^2}}} = \frac{{k_{Ft}^3}}{{3{\pi ^2}}}$ , defined via the particle densities $n_i^0$ in the absence of interactions and at zero temperature, as well as the Fermi momentum for each band $k_{Fi}$  and the total Fermi momentum $k_{Ft}$.
According to the model of the two-band system we also assume the presence of the energy shift between bands ${E_g} = \eta {E_{F2}}$, where $E_{F2}=k_{F2}^2/2m$. This implies the relations between different Fermi momentums ${k_{F1}} = {\left[ {1 - \frac{1}{{{{\left( {\eta  + 1} \right)}^{\frac{3}{2}}} + 1}}} \right]^{\frac{1}{3}}}{k_{Ft}}$ and ${k_{F2}} = {\left[ {\frac{1}{{{{\left( {\eta  + 1} \right)}^{\frac{3}{2}}} + 1}}} \right]^{\frac{1}{3}}}{k_{Ft}}$.
For regularization we use the s-wave scattering lengths for each band $a_{ii}$  defined by the low-energy limit of the two-body problem in vacuum
\begin{equation}
\label{eq6}
\frac{m}{{4\pi {a_{ii}}}} =  - \frac{1}{{{U_{ii}}}} + \sum\limits_{\bf{k}}^{{{\bf{k}}_{\bf{0}}}} {\frac{m}{{{{\bf{k}}^2}}}},
\end{equation}
with increase of the momentum cut-off $k_0$ which is much larger than the average distance between particles and $\left| k \right| \le {k_0}$.  We will show that in calculations the selection of the cut-off momentum value will not affect for the obtained results (see Appendix A and B). For the sake of simplification we redefine constants $U_{ij} = {{{\tilde U}_{ij}}{{\left( {\frac{{{k_{Ft}}}}{{{k_0}}}} \right)}^2}\frac{{{E_{Ft}}}}{n}}$ of the intraband ($i=j$) and the interband ($i \ne j$) coupling, where $E_{Ft}=k_{Ft}^2/2m$ is the total Fermi energy. From Eq. (\ref{eq6}) this yields relations in the dimensionless form between intraband coupling coefficients and scattering lengths for each band
\begin{equation}
\label{eq7}
{{\tilde U}_{11}}{\left( {\frac{{{k_{Ft}}}}{{{k_0}}}} \right)^2} = \frac{4}{3}{\left( {\frac{{{k_0}}}{{{k_{Ft}}}} - \frac{\pi }{{2{k_{F1}}{a_{11}}}}\frac{{{k_{F1}}}}{{{k_{Ft}}}}} \right)^{ - 1}}.
\end{equation}
Substituting Eq. (\ref{eq7}) to Eq. (\ref{eq3}) and performing dimensionless procedure for Eqs. (\ref{eq3}) and (\ref{eq5}) in units of the total Fermi momentum and the total Fermi energy, we get the system of equations for the energy gaps and the particle densities that will be solved numerically (see Appendix A).

Besides the energy gaps $\Delta_i$ and the particle densities $n_i$ another important characteristic of the pairing regimes through out the BCS-BEC crossover in a two-band superfluid Fermi gas is the intrapair correlation lengths of the Cooper pairs, which is determined by the expression 
\begin{equation}
\label{eq11}
\xi _{{\rm{pair,}}i}^2 =\frac{{\sum\limits_{\bf{k}} {{{\left| {{\nabla _{\bf{k}}}\left( {\frac{{1 - 2f\left( {{E_i}\left( {\bf{k}} \right)} \right)}}{{{E_i}\left( {\bf{k}} \right)}}} \right)} \right|}^2}} }}{{\sum\limits_{\bf{k}} {{{\left( {\frac{{1 - 2f\left( {{E_i}\left( {\bf{k}} \right)} \right)}}{{{E_i}\left( {\bf{k}} \right)}}} \right)}^2}} }},
\end{equation}
obtained from the pair correlation function, evaluated at a mean-field level for zero and finite temperature \cite{Palestini0}.

Differently to the paper \cite{Iskin2} where the ratio of intraband coupling constants was fixed for the investigation of the BCS-BEC crossover properties, we study a two-band system with the fixed value of scattering length for the first band, which corresponds to the BCS regime, namely $1/(k_{F1}a_{11}) = -2$, and varying scattering length for the second band $1/(k_{F2}a_{22})$ . Such a strategy allows to avoid the convergence problem and the dependence of physical quantities on the cut-off momentum value $k_0$ (see Appendix B).  During our investigations we fix the energy shift between bands $\eta = 3$, which gives $E_{\rm g}=0.75E_{\rm F1}=3E_{\rm F2}$ and  corresponding relations for Fermi momenta in each band and the total Fermi momentum $k_{\rm F1}=\left(8/9\right)^{1/ 3}k_{\rm Ft}$ and $k_{\rm F2}=\left(1/9\right)^{1/ 3}k_{\rm Ft}$. For the sake of better presentation of results and their interpretation we measure energy gaps and the chemical potential in units of the total Fermi energy $E_{Ft}$. 

\section{Results and discussion}

\subsection{Energy gaps, chemical potential and particle densities}

To provide comprehensive description of the BCS-BEC crossover properties in a two-band superfluid Fermi gas first of all we analyze the evolution of the energy gaps, the chemical potential and the particle densities at the zero temperature based on the numerical solution of Eqs. (\ref{eq3})-(\ref{eq5}). It should be noted that in principle for $T=0$  the system of Eqs. (\ref{eq3})-(\ref{eq5}) can be integrated and after long but straightforward calculations is expressed via full elliptic integrals of the first and the second kinds. These analytical calculations show that within our strategy with the fixed value of the scattering length in the first band there is no dependence on the cut-off momentum value $k_0$ for $\Delta_i$ and $\mu$ at least for the zero temperature. The same statement can be extended for the case of $T_c$ (see Appendix A).

\begin{figure*}
\includegraphics[width=0.68\columnwidth]{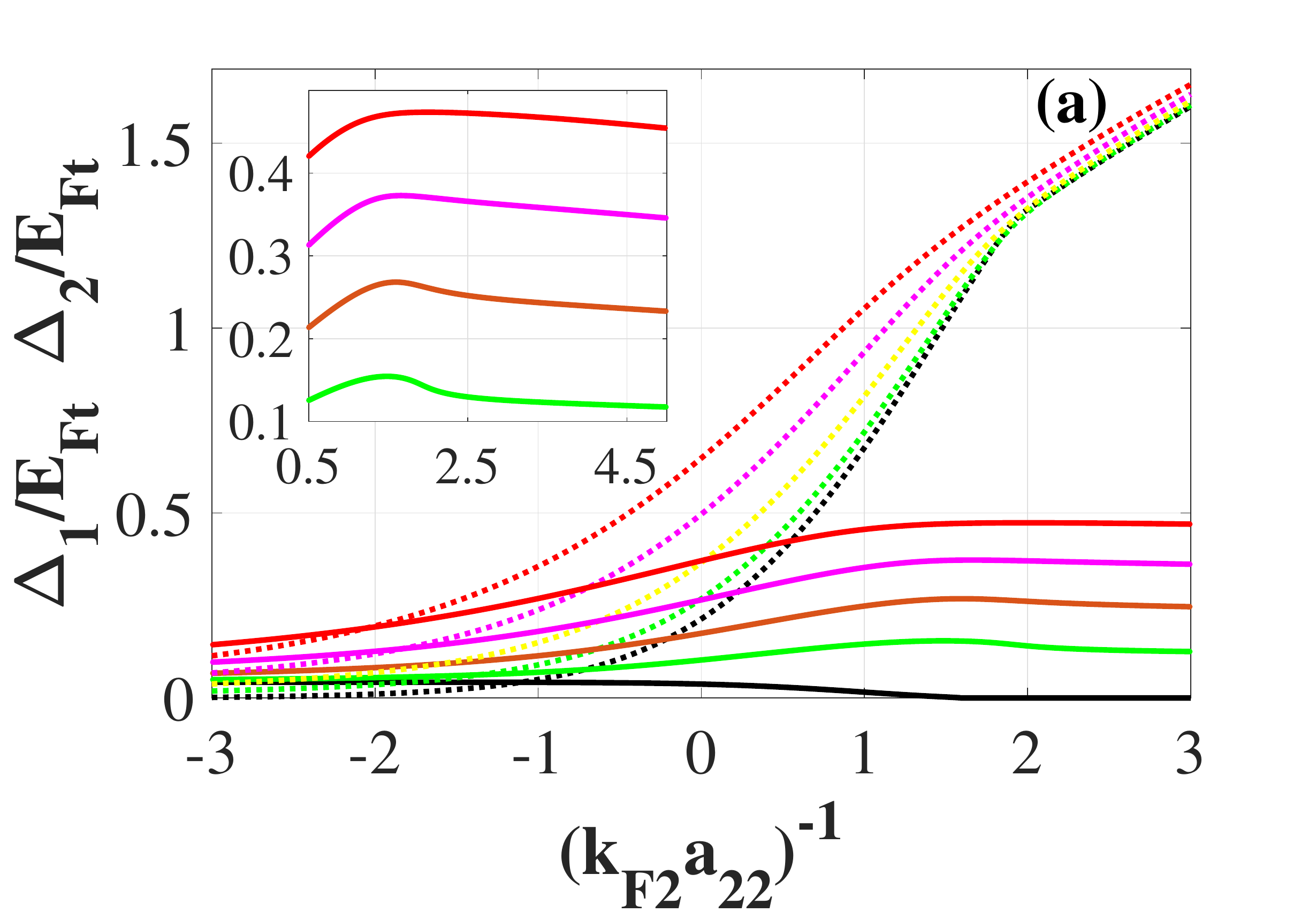}
\includegraphics[width=0.68\columnwidth]{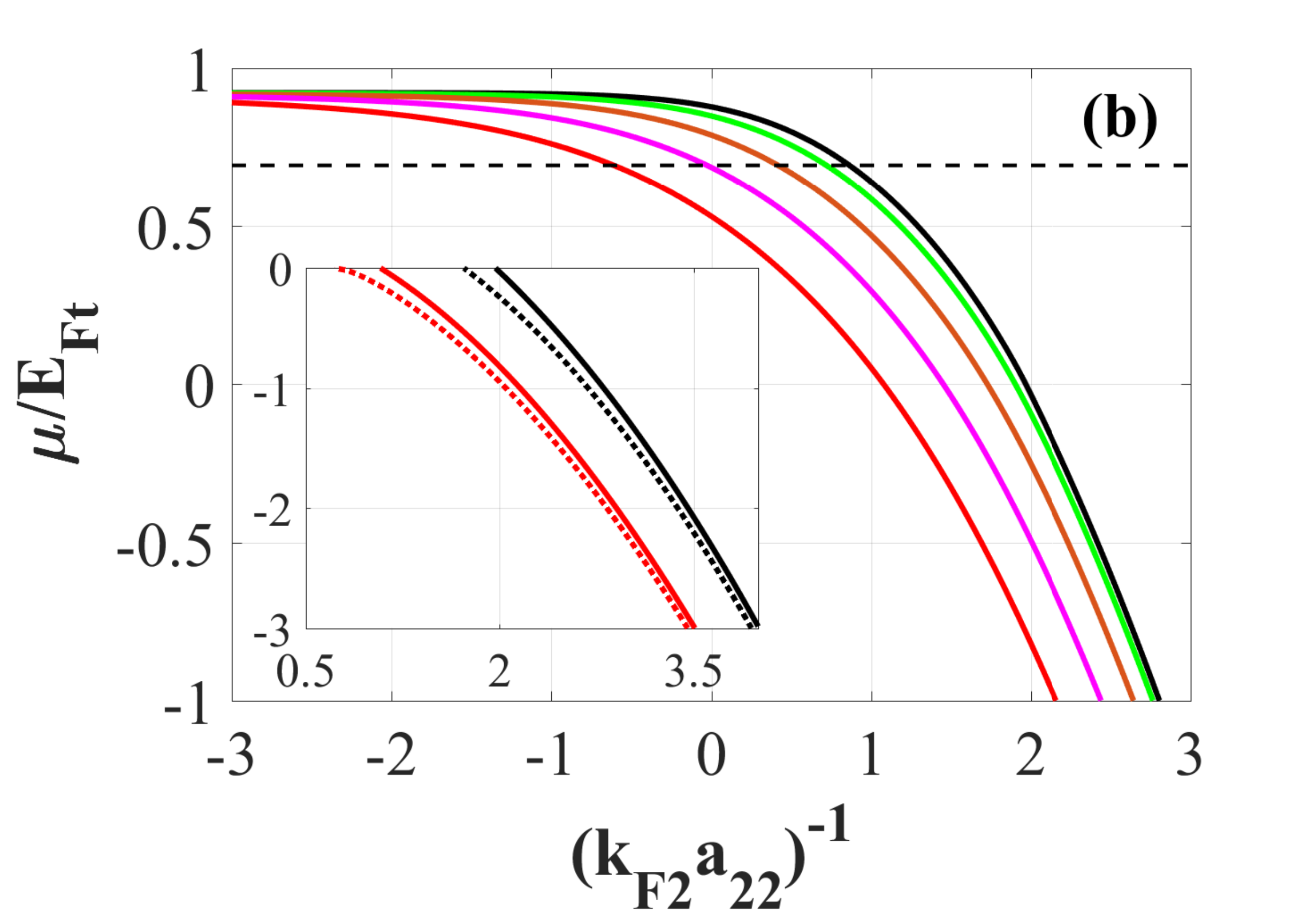}
\includegraphics[width=0.68\columnwidth]{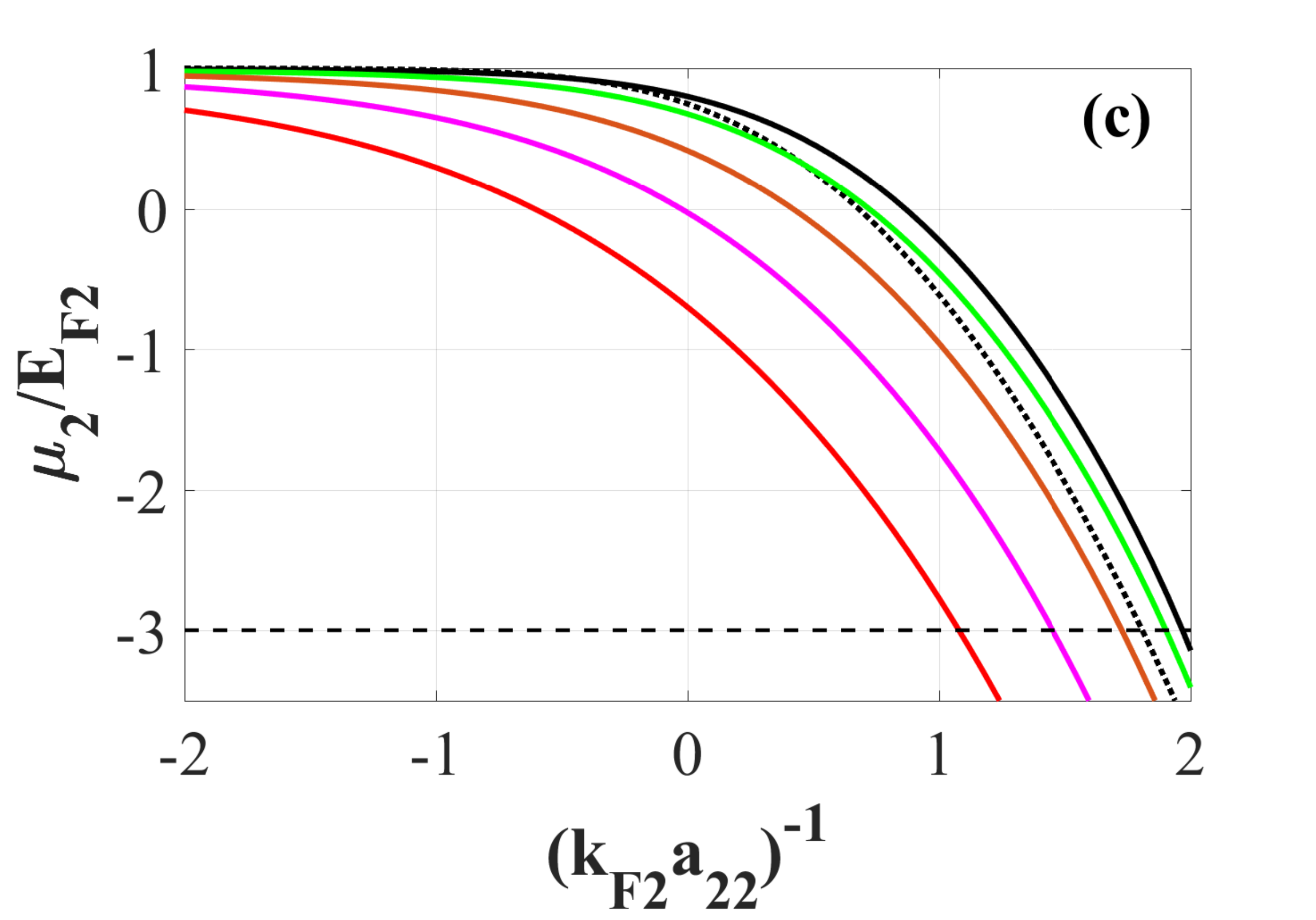}
\caption{(a) Energy gaps $\Delta_1$ (solid lines), $\Delta_2$ (dotted lines), (b) the chemical potential $\mu$ and (c) the chemical potential $\mu_2 = \mu - E_g$ in units of $E_{F2}$ at zero temperature as a function of $(k_{\rm F2}a_{22})^{-1}$ for different interband couplings  $\tilde{U}_{12} = 0$ (black line), $\tilde{U}_{12} = 0.5$ (green line), $\tilde{U}_{12} = 1$ (brown line), $\tilde{U}_{12} = 1.5$ (magenta line),  $\tilde{U}_{12} = 2$ (red line) with the fixed value of the scattering length in the first band $(k_{\rm F1}a_{11})^{-1} = -2$. The presence of the hump on energy gap  $\Delta_1$ dependencies for different interband interaction coefficients is shown in inset.
Dashed black line in (b) and (c) corresponds to the energy shift $E_g$ between bands in units of $E_{Ft}$ and $E_{F2}$ respectively. Dotted black line in (c) is the chemical potential of a single-band superfluid Fermi gas. Inset in (b) shows the comparison between the chemical potential (solid lines) and the half of the binding energy $-E_b/2$ (dotted lines) dependences in the strong-coupling limit. }
\label{fig2}
\end{figure*}

We found that in a system with vanishing interband interaction the BCS-BEC crossover is characterized by the full suppression of the first energy gap in the BEC limit and the presence of the kink on the second gap dependence at $(k_{\rm F2}a_{22})^{-1}  \approx 2 $ (Fig. 2a). 
The interband coupling smooths out the kink of the second gap and leads to the activation of the first gap in the BEC limit. Despite that Fig. 2a shows the almost constant character of  $\Delta_1$ dependence for the interval ${\left( {{k_{F2}}{a_{22}}} \right)^{ - 1}} \in \left[ {-3;3} \right]$  according to our numerical analysis we observe very slow growth of the first gap starting from the nonzero value of $\Delta_1^{(0)}  \approx 0.043$ in the BCS limit and very slow decrease of  $\Delta_1 (1/k_{F2}a_{22})$ in the BEC limit. Moreover the behavior of $\Delta_1 (1/k_{F2}a_{22})$ always has non-monotonic dependence with the very tiny hump in the BCS limit in the case of vanishing interband interaction. With the further increasing of the interband coupling this hump becomes more pronounced and is shifted to the BEC limit (see inset in Fig. 2a). Note that for weak interband coupling the energy gap in the first band is exponentially suppressed when the cold band is almost depleted. The overall non-monotonic behavior of $\Delta_1$ as a function of $(k_{\rm F2}a_{22})^{-1}$ indicates a first regime of weak to intermediate coupling in the hot band in which $\Delta_1$ increases because of the effective attraction generated by the interband interaction able to transfer attractive pairing from the hot to the cold band. On the other hand, when the coupling in the hot band becomes very strong, the depletion of the cold band starts to dominate, causing a decrease in $\Delta_1$ and the presence of the hump in $\Delta_1$ is the result of this interplay. 

In turn the chemical potential of a two-band superfluid Fermi gas decreases slower in comparison with a single-band counterpart even in the presence of the interband interaction (Fig. 2b). It is important to note that the single-band case differs from the two-band one  for $U_{12} = 0$, because a particle transfer between the two bands occurs due to the additive structure of the density equation in Eq. (\ref{eq5}). 

Based on the two-body Schr\"{o}dinger equation we calculate the dependence of the two-body binding energy $E_b$  for different interband couplings (inset in fig. 2b). One can see that when $U_{12}$ increases,  $E_b$ also increases, and the chemical potential of the system tends to the BEC limit $-$$E_b/2$.

\begin{figure}
\includegraphics[width=1\columnwidth]{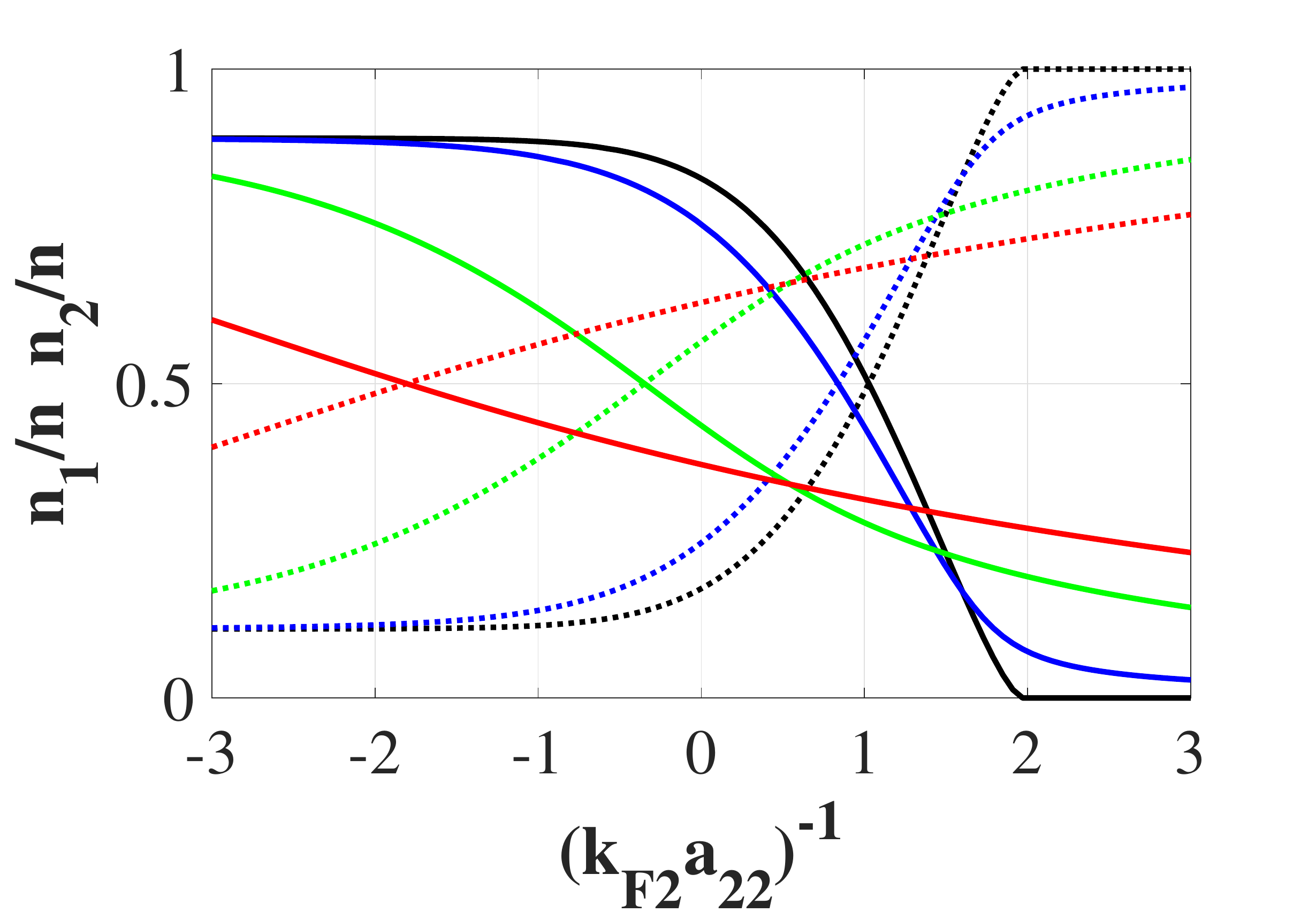}

\caption{Distribution of particle densities in each band $n_1/n$ (solid lines) and  $n_2/n$ (dotted lines) normalized on the total particle densities as a function of $(k_{\rm F2}a_{22})^{-1}$ for different interband couplings  $\tilde{U}_{12} = 0$ (black line), $\tilde{U}_{12} = 1$ (blue line), $\tilde{U}_{12} = 3$ (green line) and $\tilde{U}_{12} = 5$ (red line) with the fixed value $(k_{\rm F1}a_{11})^{-1} = -2$ and  at $T = 0$.}
\label{fig3}
\end{figure}

In Figure 2c we report the behavior of $\mu_2 = \mu - E_g$ normalized to $E_{F2}$ as a function of $(k_{\rm F2}a_{22})^{-1}$ to compare with the single-band result. For vanishing or weak interband coupling ($\tilde{U}_{12} < 1$), the chemical potential can be larger than for a single-band case. This indicates Pauli-blocking effects due to the cold states as already obtained in the vicinity of the critical temperature by a Nozi\`{e}res-Schmitt-Rink approach in Ref. \onlinecite{Tajima}.

For $\tilde U_{12}=\tilde U_{21}=0$ the full suppression of $\Delta_1$ is connected with the full redistribution of particles between bands (Fig. 3). Even though the density equation Eq.  (\ref{eq5}) couples two condensates, the first deep band remains in the BCS regime and till unitarity the particle distribution among two bands is not important for the system. The increasing of the interband interaction leads to particles equalizing in the weak-coupling regime and retardation of the distribution process between bands in the strong-coupling limit. 

\begin{figure}
\includegraphics[width=0.49\columnwidth]{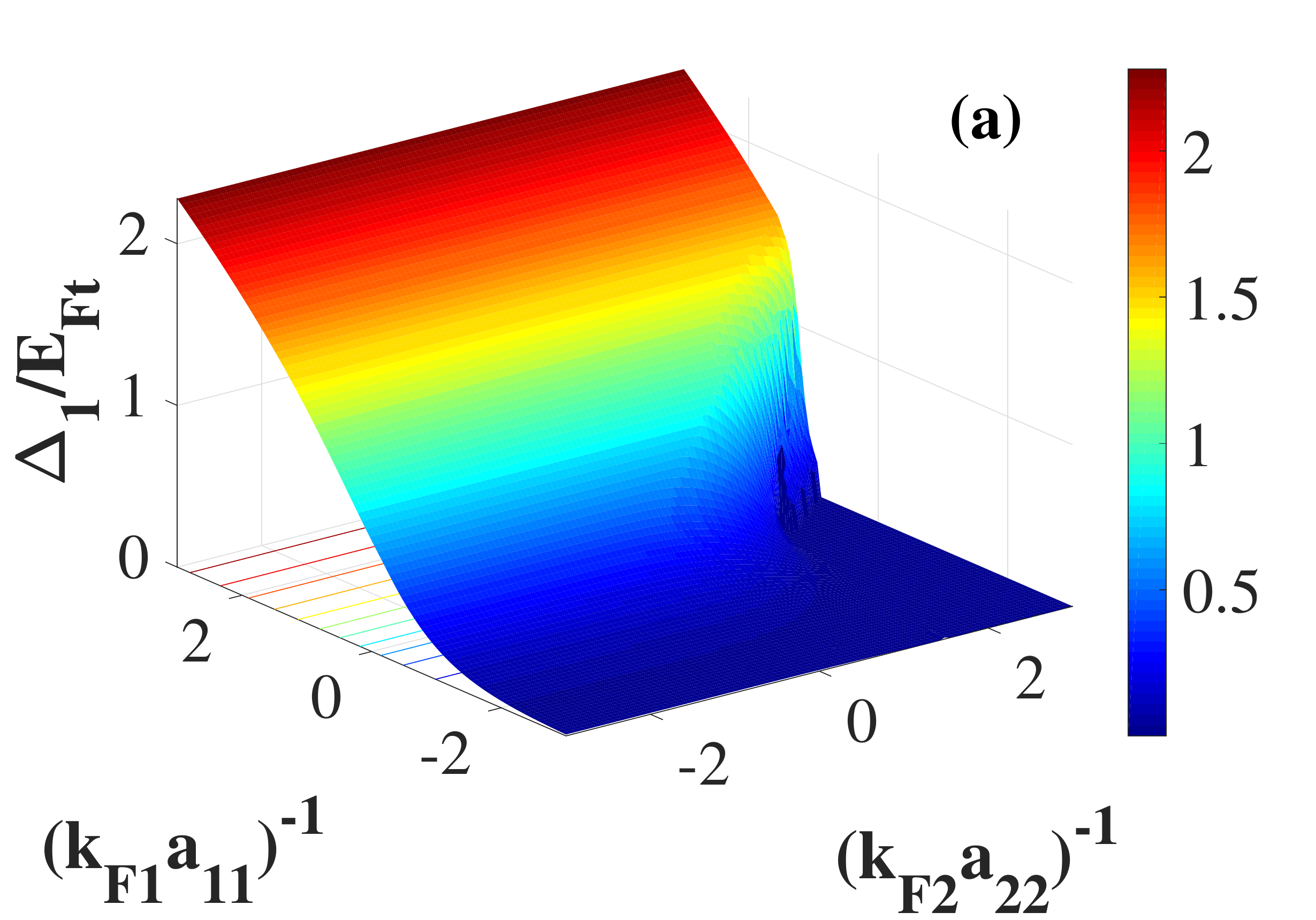}
\includegraphics[width=0.49\columnwidth]{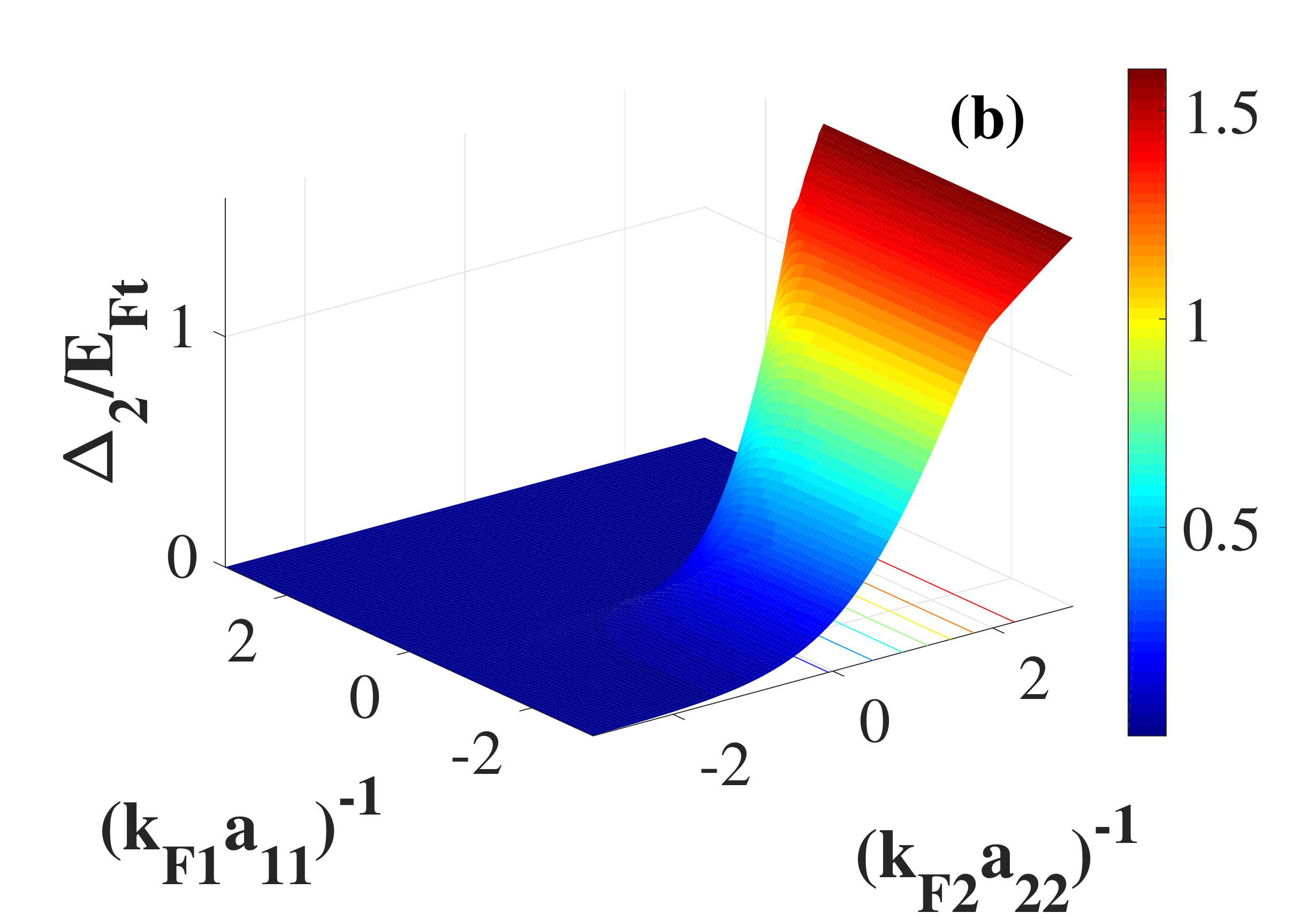}
\includegraphics[width=0.49\columnwidth]{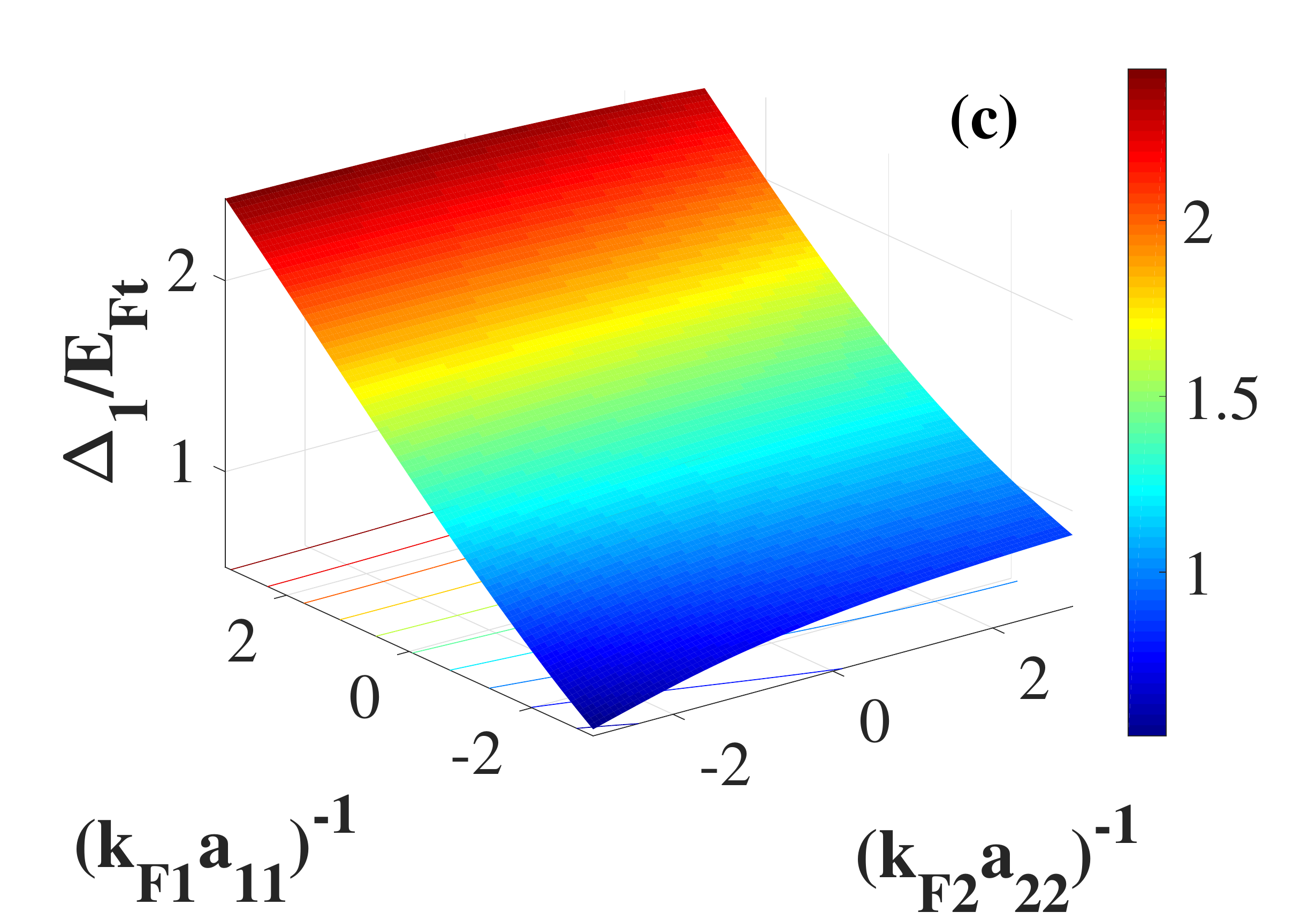}
\includegraphics[width=0.49\columnwidth]{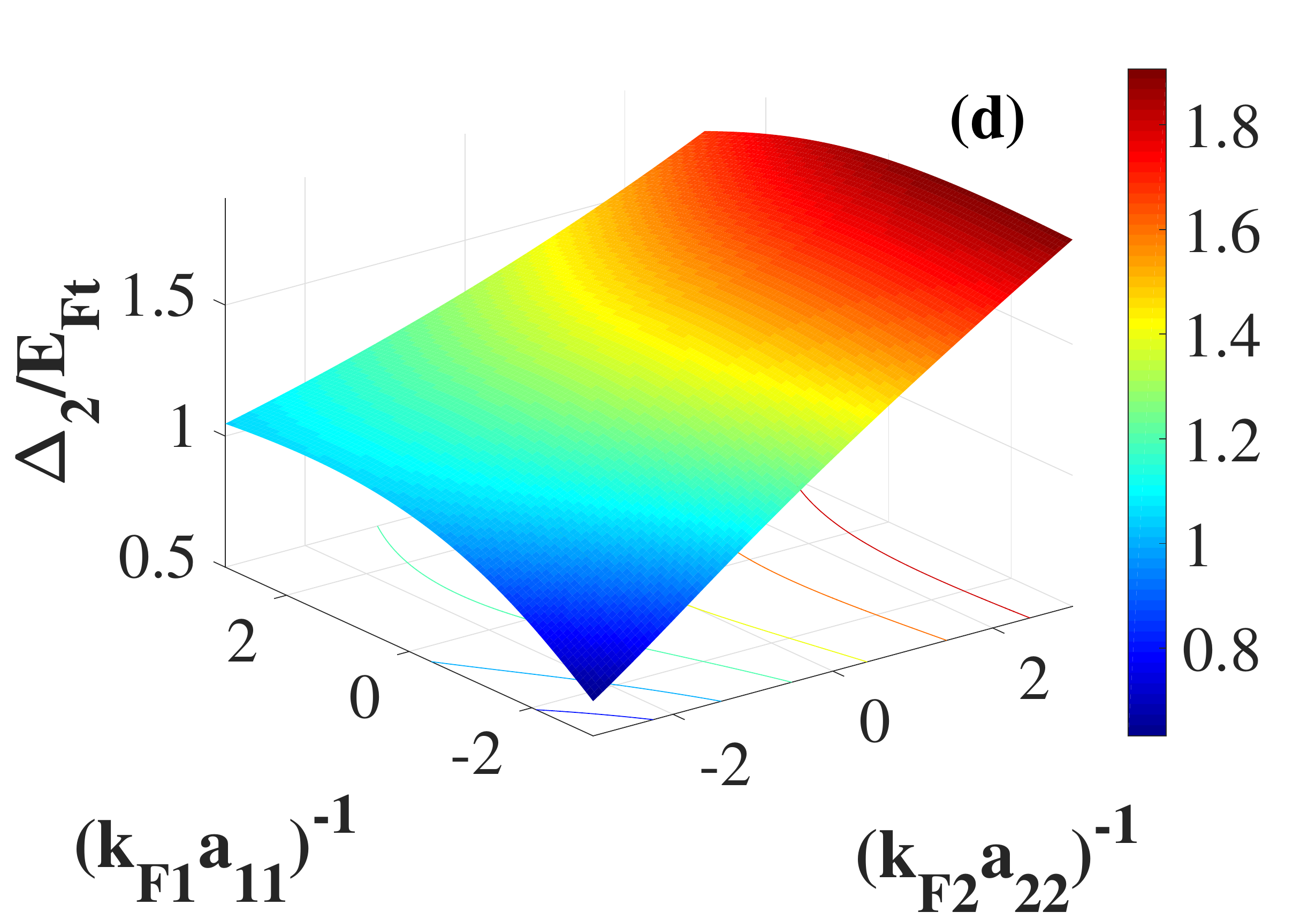}
\caption{Evolution of energy gaps $\Delta_1$ (a), (c) and $\Delta_2$ (b), (d) at zero temperature vs. scattering lengths for fermions in the first and second bands for different interband couplings: $\tilde{U}_{12} = 0$ (a), (b) and strongly interacting bands $\tilde{U}_{12} = 5$ (c), (d).}
\label{fig4}
\end{figure}

Until now we have investigated the characteristics of the BCS-BEC crossover in a two-band fermionic system with the fixed value of intraband coupling in the first band $(k_{\rm F1}a_{11})^{-1} = -2$. To understand full properties of  the BCS-BEC crossover for this system at $T=0$, we consider the behavior of the energy gaps and the particle densities varying $(k_{\rm F1}a_{11})^{-1}$. First of all as we can see from Fig. 4a and Fig. 4b weak intraband coupling in the first band and the strong in the second one, namely when $(k_{\rm F1}a_{11})^{-1} < 0 $  and $(k_{\rm F2}a_{22})^{-1} > 2 $ together with the vanishing interband interaction transform the two-band system into the single-band one with the fully suppressed first gap. The single-band scenario is realized also for $(k_{\rm F1}a_{11})^{-1} > 0 $ and for the entire interval of values  $(k_{\rm F2}a_{22})^{-1}$ with the full suppression of the second gap. Increasing of the interband interaction extends the region and broadening the borders, where the two-band configuration is preserved (fig. 4c and fig. 4d). 
\begin{figure}
\includegraphics[width=0.49\columnwidth]{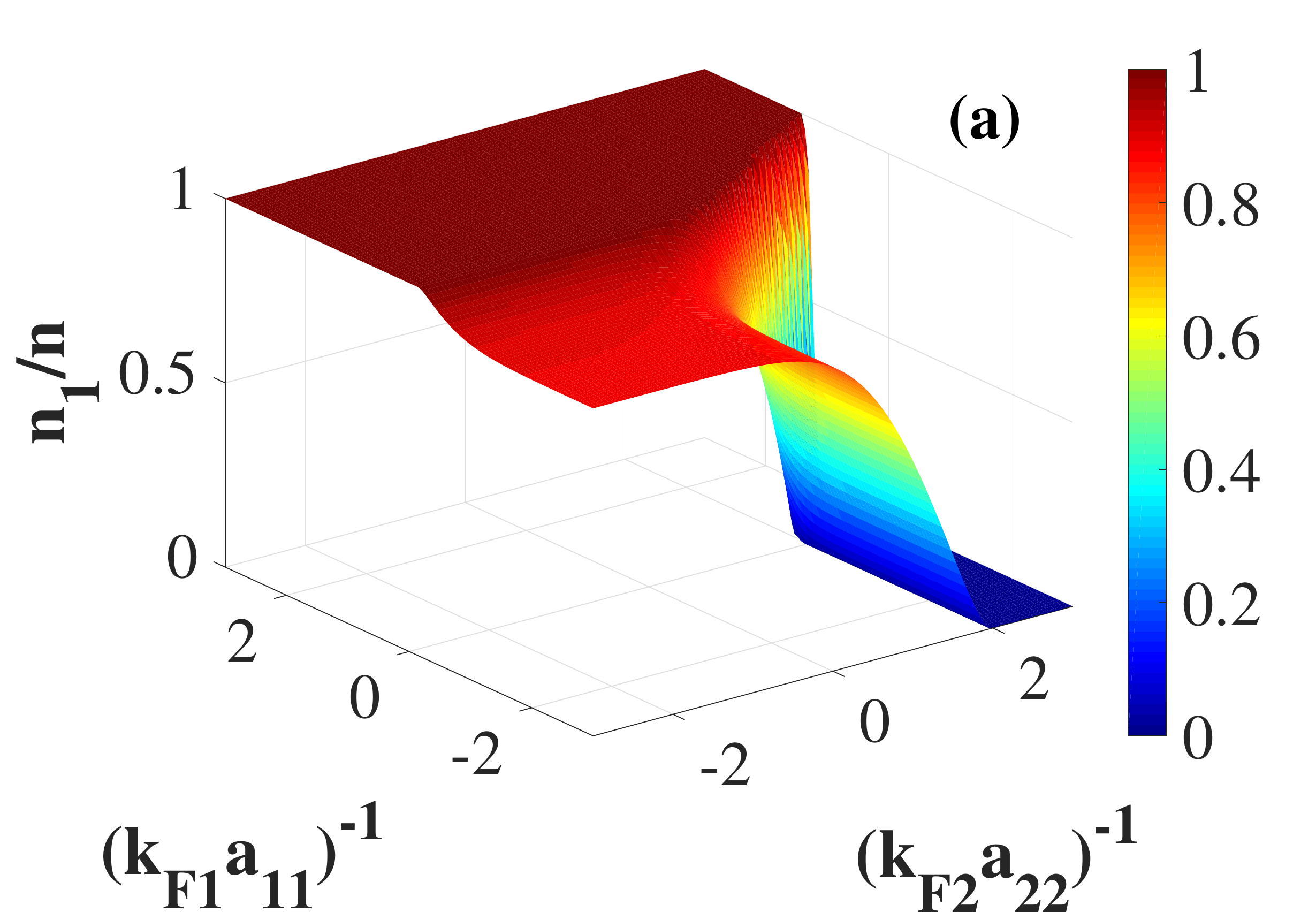}
\includegraphics[width=0.49\columnwidth]{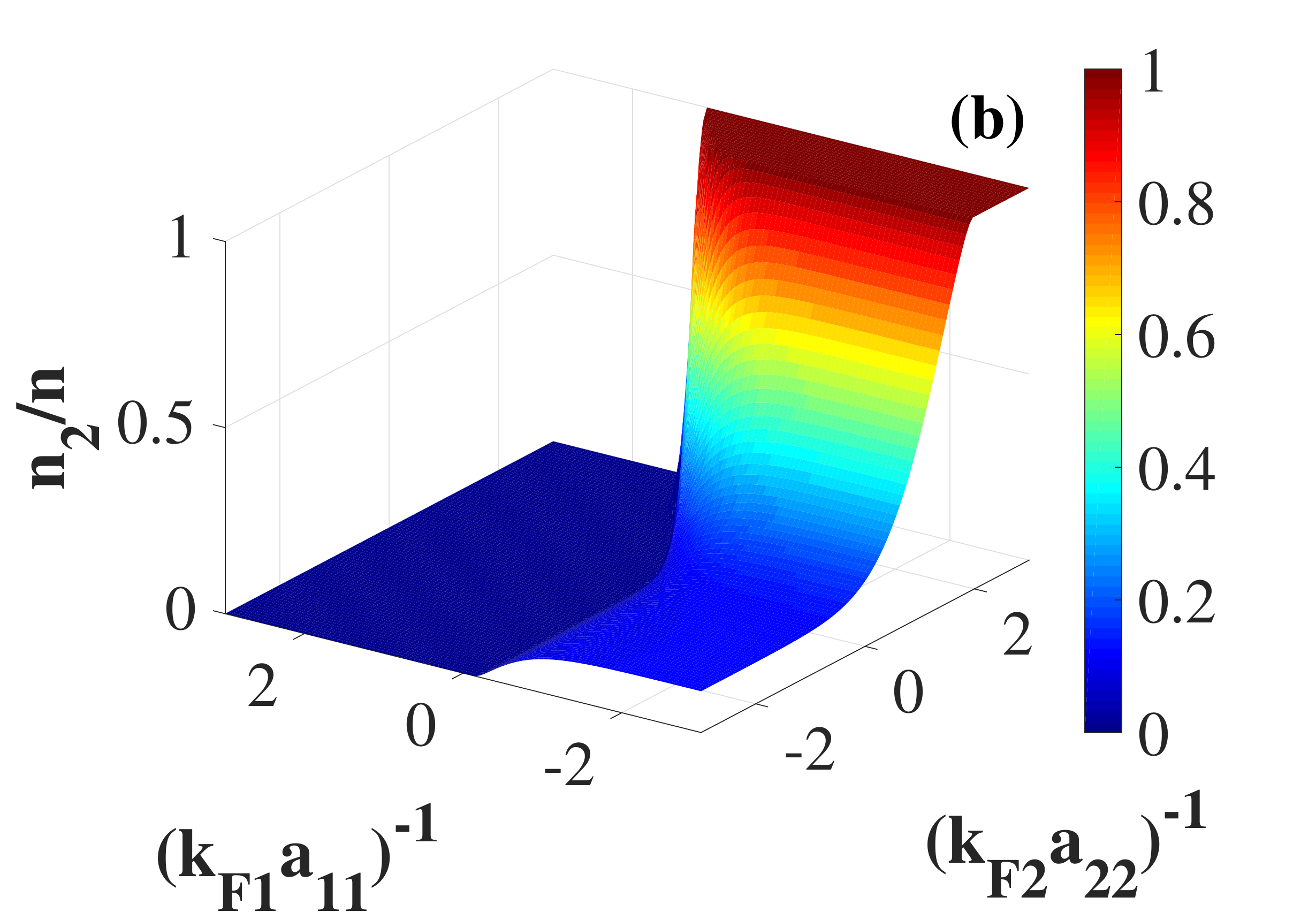}
\includegraphics[width=0.49\columnwidth]{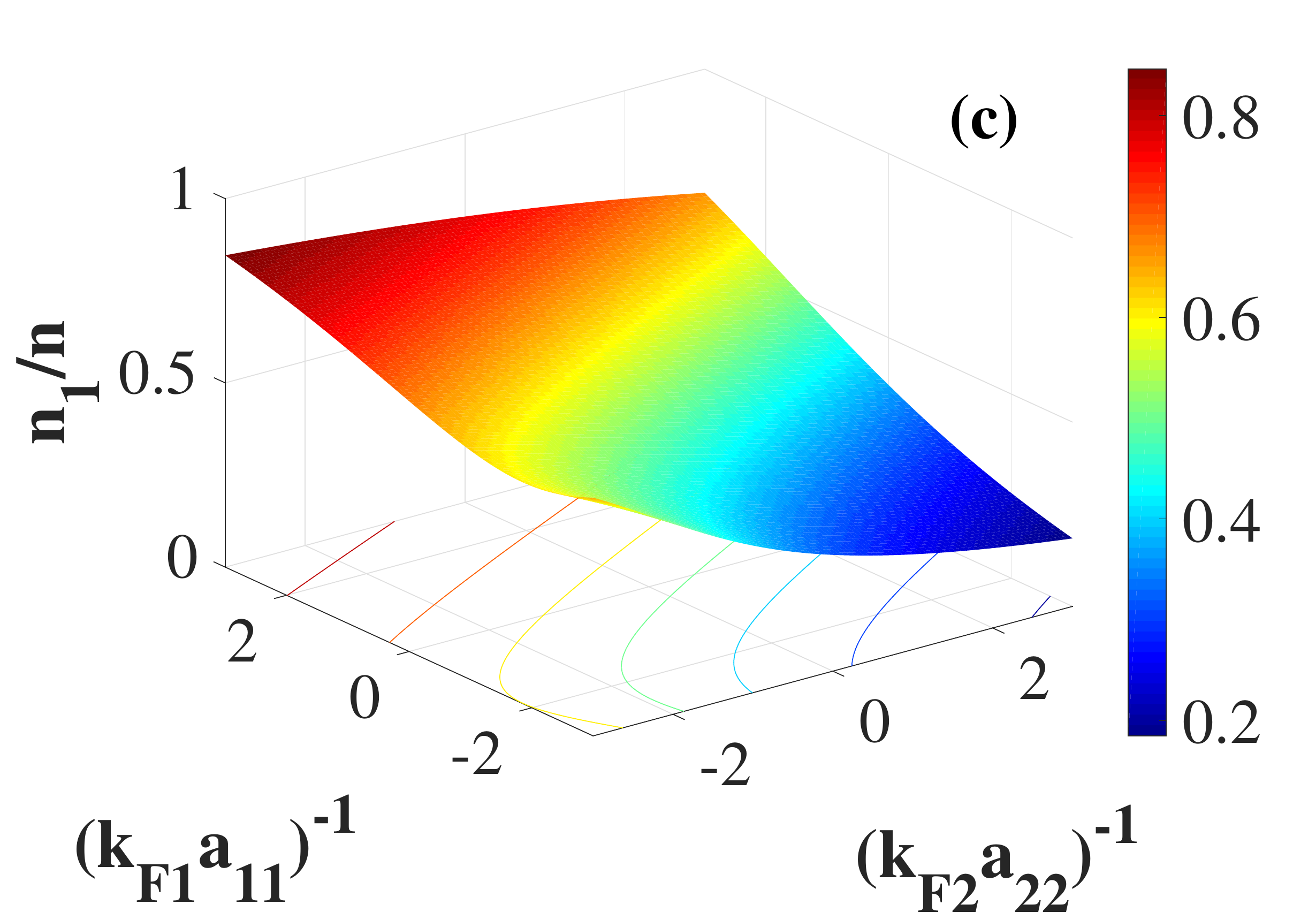}
\includegraphics[width=0.49\columnwidth]{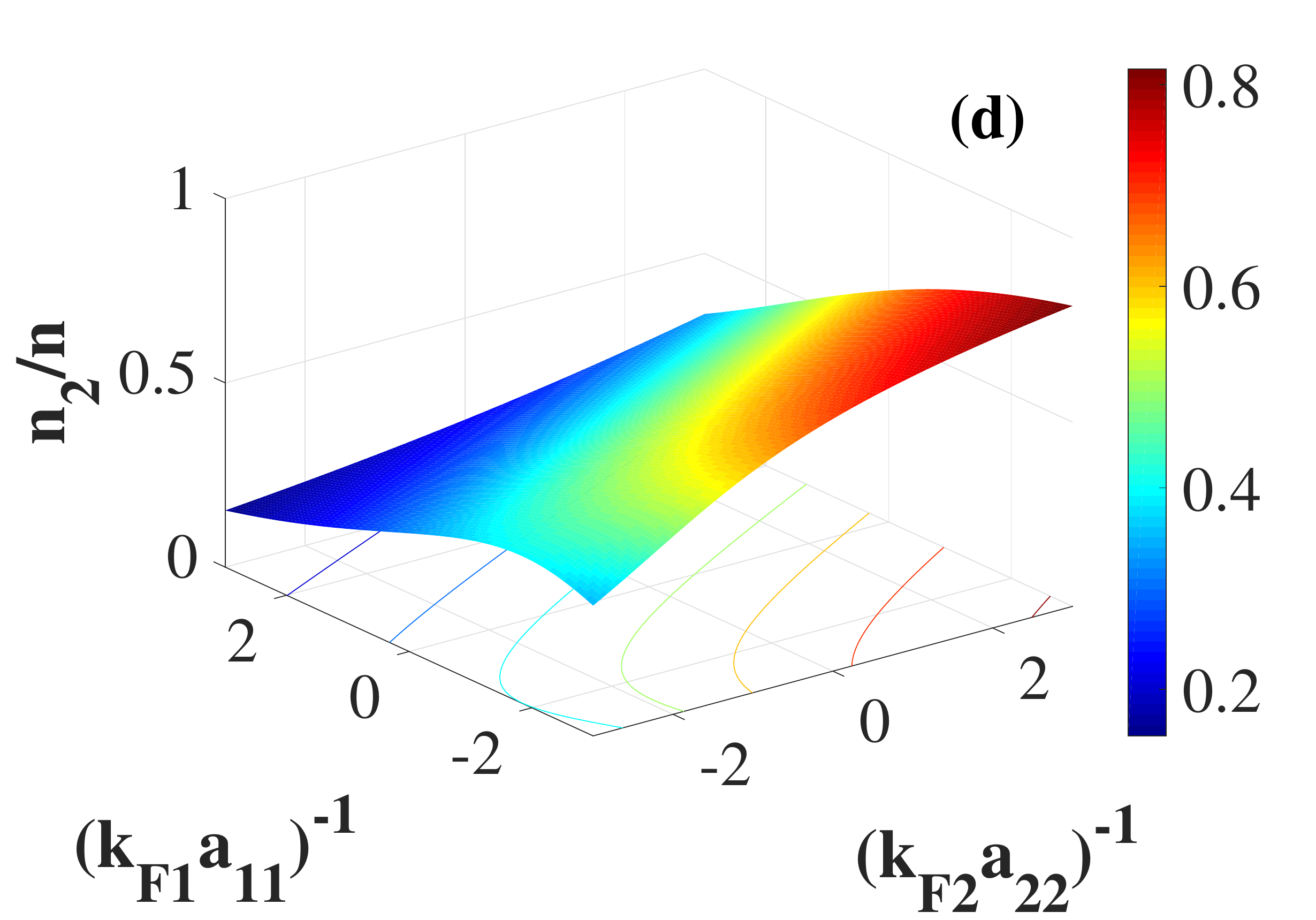}
\caption{Three-dimension plots of normalized particle densities in the first (a), (c) and in the second band (b), (d)  as a function of $(k_{\rm F1}a_{11})^{-1}$ and $(k_{\rm F2}a_{22})^{-1}$  for different values of interband interaction coefficients  $\tilde{U}_{12} = 0$ (a), (b) and $\tilde{U}_{12} = 5$ (c), (d) at $T = 0$.}
\label{fig5}
\end{figure}

It is worth noting that a similar transition from single-condensate to two-condensate superconductivity was revealed experimentally in the ${\rm{LaAl}}{{\rm{O}}_{\rm{3}}}{\rm{/SrTi}}{{\rm{O}}_{\rm{3}}}$ interface driven by electrostatic doping  \cite{Singh1}. It was found that in such a heterostructure the superconducting gap in the first band is suppressed while the second band is populated. Within our approach, we speculate that these results can be interpreted as the transition from the BEC to the BCS regime of a two-band superfluid system close to a Lifshitz transition with vanishing interband interaction. Despite the fact that heterostructures ${\rm{LaAl}}{{\rm{O}}_{\rm{3}}}{\rm{/SrTi}}{{\rm{O}}_{\rm{3}}}$ represent the two-dimensional electron liquid in the interface, some theoretical models argue the importance of the three-dimensional bands for the explanation of 2D superconductivity in these systems  \cite {Fernandes1}.

Our results are confirmed also by the evolution of the particle densities, where as it can be seen from Fig. 5a and Fig. 5b there is no transfer of particles for vanishing interband interaction for $(k_{\rm F1}a_{11})^{-1} > 0 $ and all particles are concentrated in the deeper band, while in the opposite case for $(k_{\rm F1}a_{11})^{-1} < 0$ and $(k_{\rm F2}a_{22})^{-1} > 2$ the bands change places: all particles migrate to the shallow band. Figures 5c and 5d show that with the increasing of the interband coupling populations in each band begin to equalize.

\subsection{Intrapair correlation lengths}

\begin{figure*}
\includegraphics[width=0.95\columnwidth]{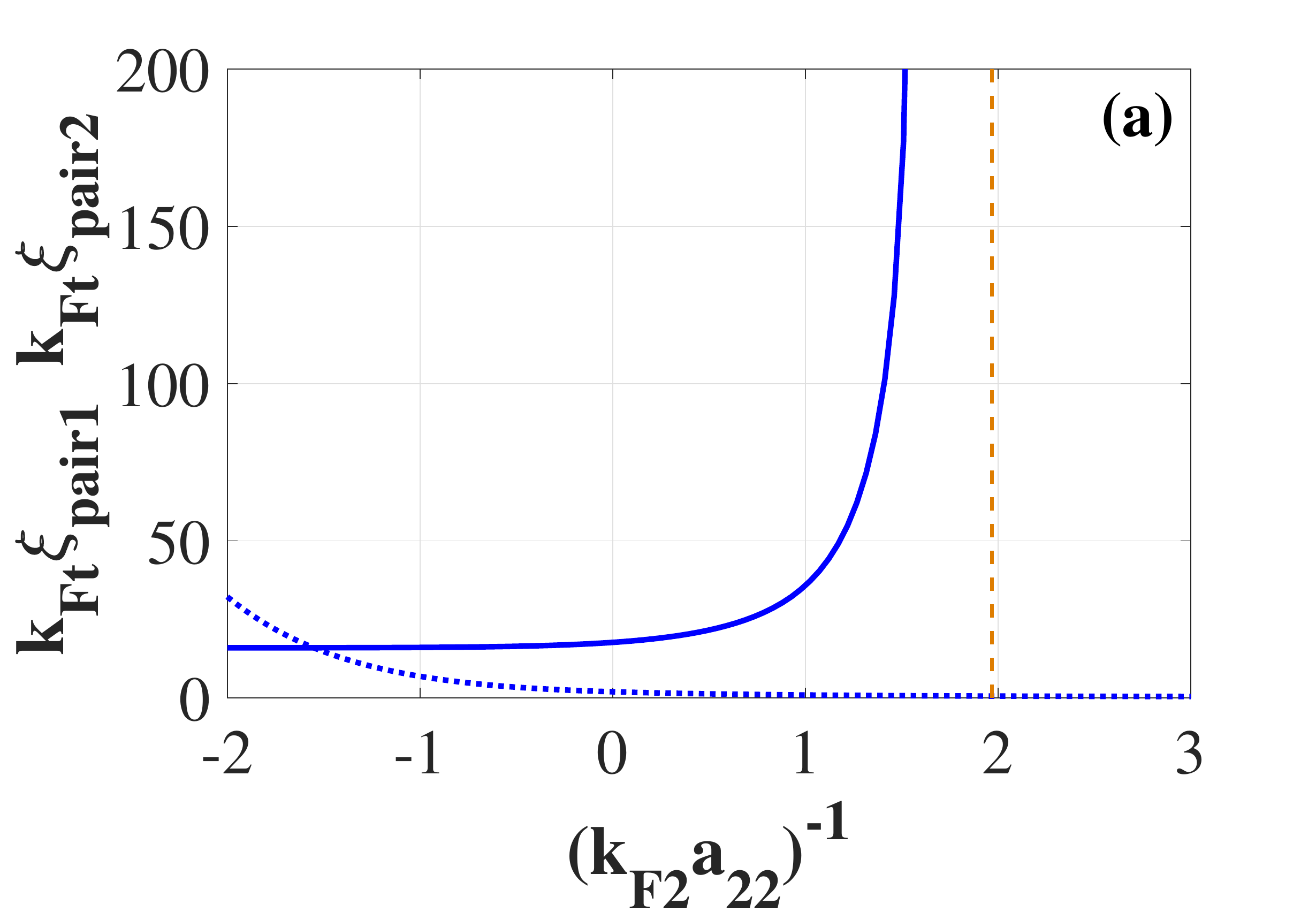}
\includegraphics[width=0.95\columnwidth]{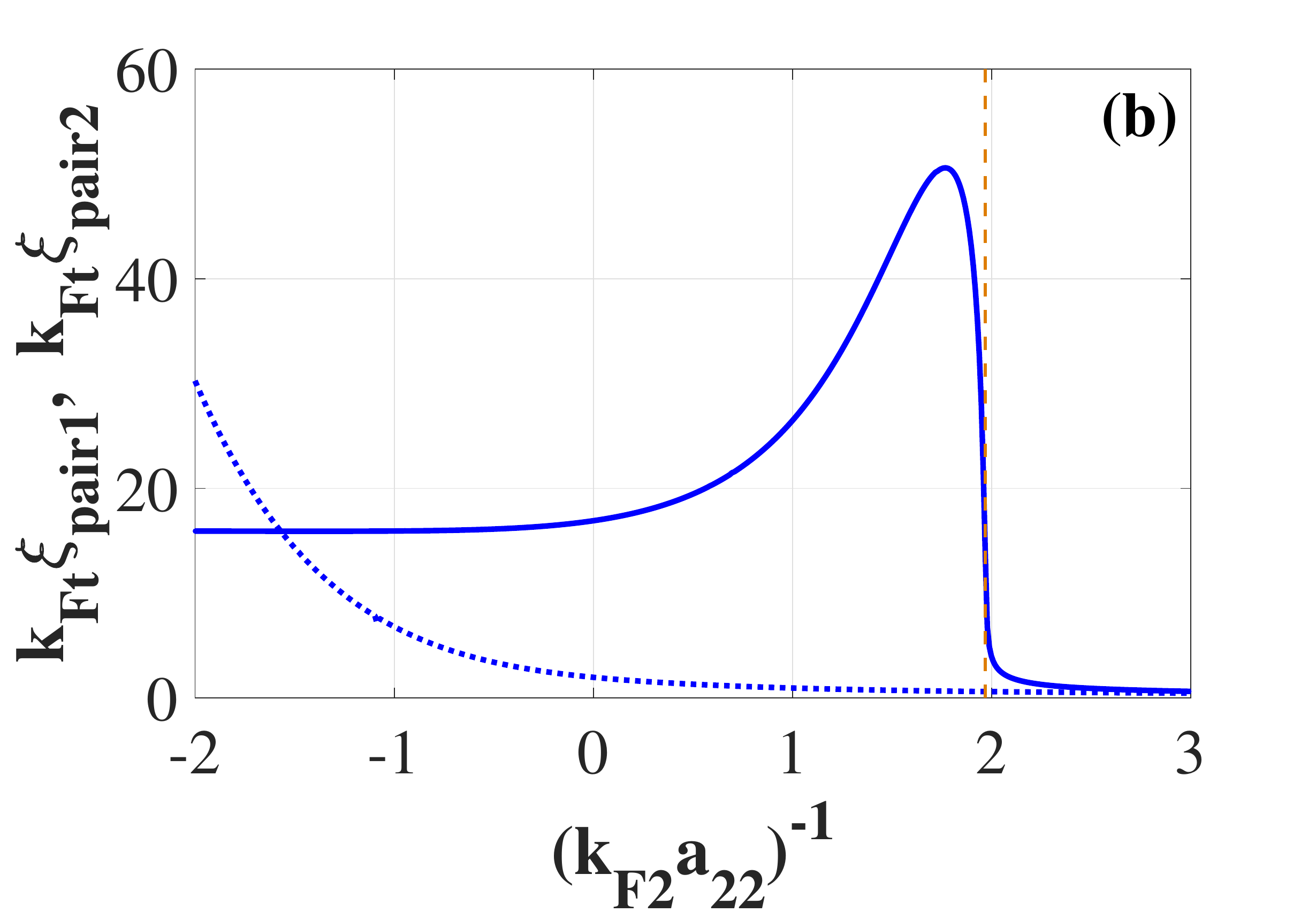}
\includegraphics[width=0.95\columnwidth]{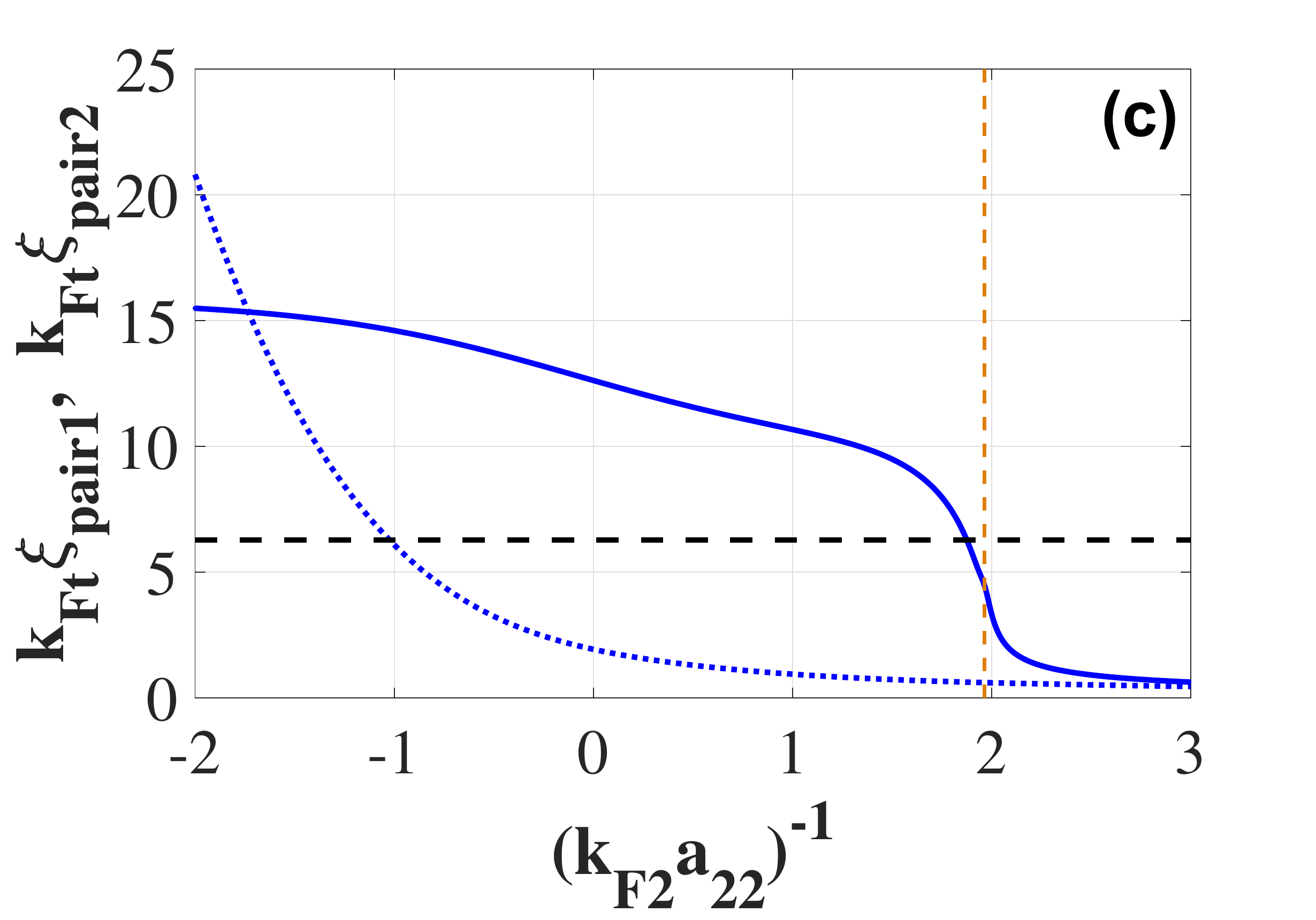}
\includegraphics[width=0.95\columnwidth]{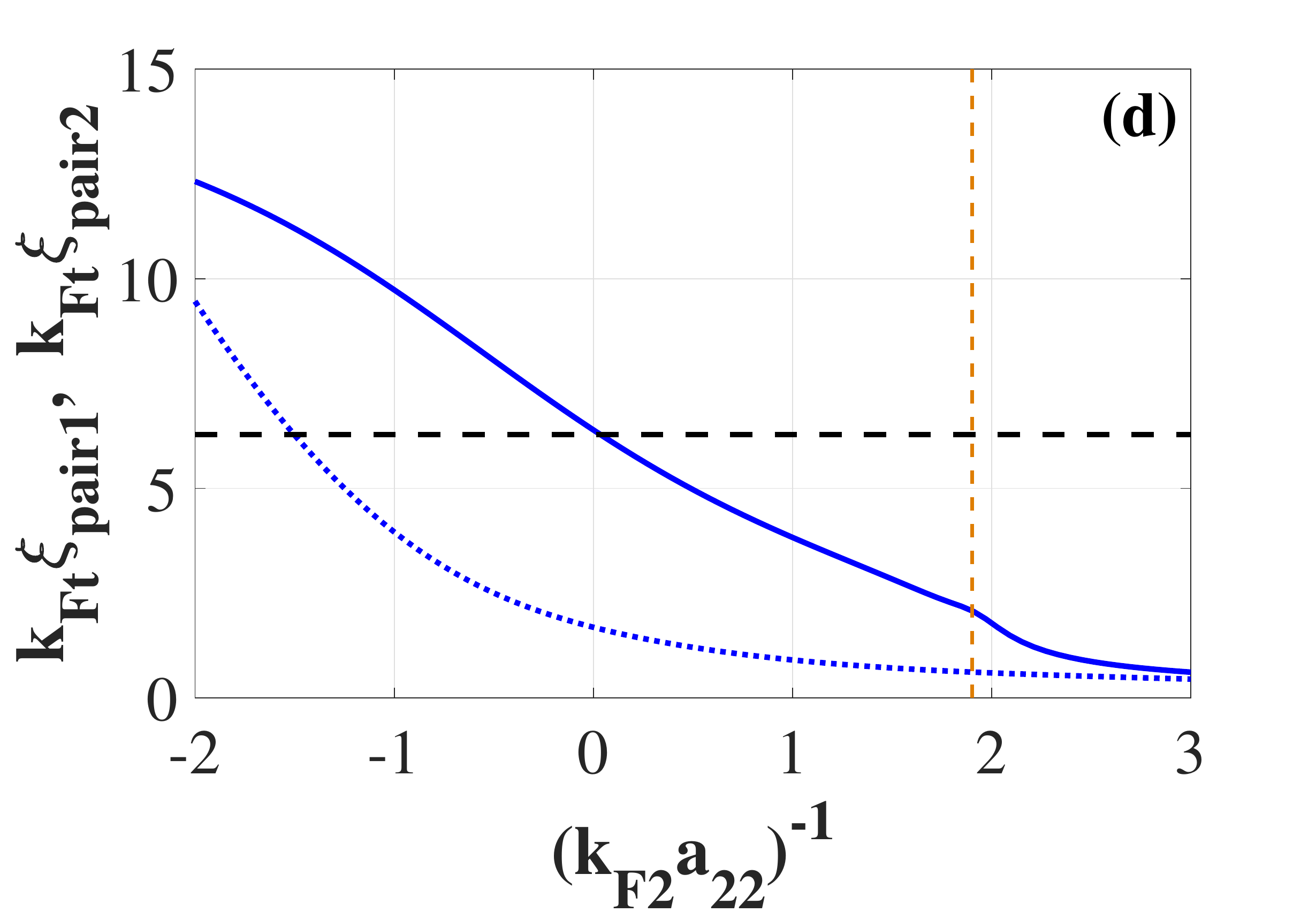}
\caption{Intrapair correlation lengths  $\xi_{\tt{pair}1}$ and  $\xi_{\tt{pair}2}$ for the first (solid lines) and the second band (dotted lines) correspondingly at zero temperature as a function of $(k_{\rm F2}a_{22})^{-1}$ for different interband couplings strengths  $\tilde{U}_{12} = 0$ (a), $\tilde{U}_{12} = 0.01$ (b), $\tilde{U}_{12} = 0.1$ (c) and $\tilde{U}_{12} = 0.5$ (d) in the case of the fixed value of $(k_{\rm F1}a_{11})^{-1} = -2$. Dashed black lines   in (c) and (d) correspond to $k_F\xi_{\tt{pair}} = 2\pi$ and delimit the BCS-BEC crossover regime (due to large values of $\xi_{\tt{pair}1}$ and  $\xi_{\tt{pair}2}$ in panels (a) and (b) we did not plot $k_F\xi_{\tt{pair}} = 2\pi$). Orange dashed lines defines the value of the coupling strength in the second band for which the chemical potential of the system equals to zero (see fig. 2b), namely $(k_{\rm F2}a_{22})^{-1} \approx 1.968$ in (a) and (b), $(k_{\rm F2}a_{22})^{-1} \approx 1.965$ in (c) and $(k_{\rm F2}a_{22})^{-1} \approx 1.903$.}
\label{fig6}
\end{figure*}

Another characteristic for the description of the crossover from Cooper-pair superconductivity to Bose-Einstein condensation of bound pairs of fermions is the intrapair correlation length that is defined by Eq. (\ref{eq11}). For a single-band system it was shown earlier that there is a universal material-independent criterion $k_F\xi$ to follow the evolution of the BCS-BEC crossover \cite{Pistolesi}. Based on Eq. (\ref{eq11}) and the definition from the paper Ref. \onlinecite{Pistolesi}, we investigate the intrapair correlation lengths for each band  $\xi_{\tt{pair}1}$ and  $\xi_{\tt{pair}2}$ as a function of the scattering length in the second band $(k_{\rm F2}a_{22})^{-1}$.
 
In the case of vanishing interband interaction  $\xi_{\tt{pair}2}$  dependence has a conventional behavior as in the single-band case, whereas  $\xi_{\tt{pair}1}$ is almost constant until the unitarity point and undergoes essential discontinuity at $(k_{\rm F2}a_{22})^{-1} \approx 2$  (Fig. 6a). The origin of this discontinuity can be understood from the definition of  $\xi_{\tt{pair}1}$ after straightforward integration of Eq. (\ref{eq11}). The obtained expression diverges when the energy gap in the first band $\Delta_1 = 0$ is fully suppressed (Fig. 2a).

By increasing the interband coupling this discontinuity is removed. For very weak interaction between bands a sharp peak on a dependence $ \xi_{\tt{pair}1} (1/k_{\rm F2}a_{22})$ is obtained (Fig. 6b). Contrary to expectations that strong-coupling limit will suppress gradually the intrapair correlation length in the first band, we observe a non-monotonic dependence and the significant amplification of  $\xi_{\tt{pair}1}$ in the BEC regime. From the physical point of view such results point out the formation of giant Cooper pairs in the first band with bosonic pairs in the second band. This coexistence of BCS and BEC condensates stems from the weak interband coupling, where the cold band serves as an almost independent reservoir of Cooper pairs. Thus, a two-band superfluid system is described as the continuous transformations from two different BCS condensates to a system where giant Cooper pairs and bosonic condensate coexists and then finally to the mixture of two BEC condensates (Fig. 7a). The crossover in the cold band above discussed can be also interpreted as a density-induced BCS-BEC crossover \cite{Andrenacci1999}, when the density $n_1$ is tuned by the coupling in the hot band. 

For larger values of $\tilde U_{12}>0.063$ and for $(k_{\rm F1}a_{11})^{-1}=-2$, we observe gradually disappearance of this peak and the transition to the conventional single-band behavior of the intrapair correlation length for the first band (Fig. 6c and d). The behavior of   $\xi_{\tt{pair}2}$ on the qualitative level maintains dependence as for the single-band counterpart. For very strong interband interaction, dependences of intrapair correlation lengths in each band are the same.

Increasing of the intraband coupling in the first band for the given strength of interband interaction leads to the reduction of the peak. Nevertheless, for very weak interband coupling, the effect of the intrapair correlation length amplification can be preserved even for $(k_{\rm F1}a_{11})^{-1}=-0.5$  (Fig. 7a). 

Using the criterion of strongly overlapping Cooper pairs for the single-band system $k_F \xi_{\tt{pair}}>2\pi$ we can extract interesting feature of the BCS-BEC crossover  in a weakly interacting two-band Fermi gas. 
In particular, when the value of $(k_{\rm F1}a_{11})^{-1} < 0$ we have rich picture of the BCS-BEC crossover evolution (Fig. 7a). Initially for $1/(k_{\rm F2}a_{22}) \ll -1$ there is a mixture of two Cooper pairs condensates. Then  with the increasing of the intraband coupling towards to the strong-coupling regime formation of giant Cooper pairs in the first band occurs. Such pairs coexist with the BEC condensate from the second band. In the extremely strong coupling limit  $1/(k_{\rm F2}a_{22})  \gg -1$ we observe transition from giant Cooper pairs and BEC molecules into the two bosonic condensates with coinciding intrapair correlation lengths. 

\begin{figure}
\includegraphics[width=0.49\textwidth]{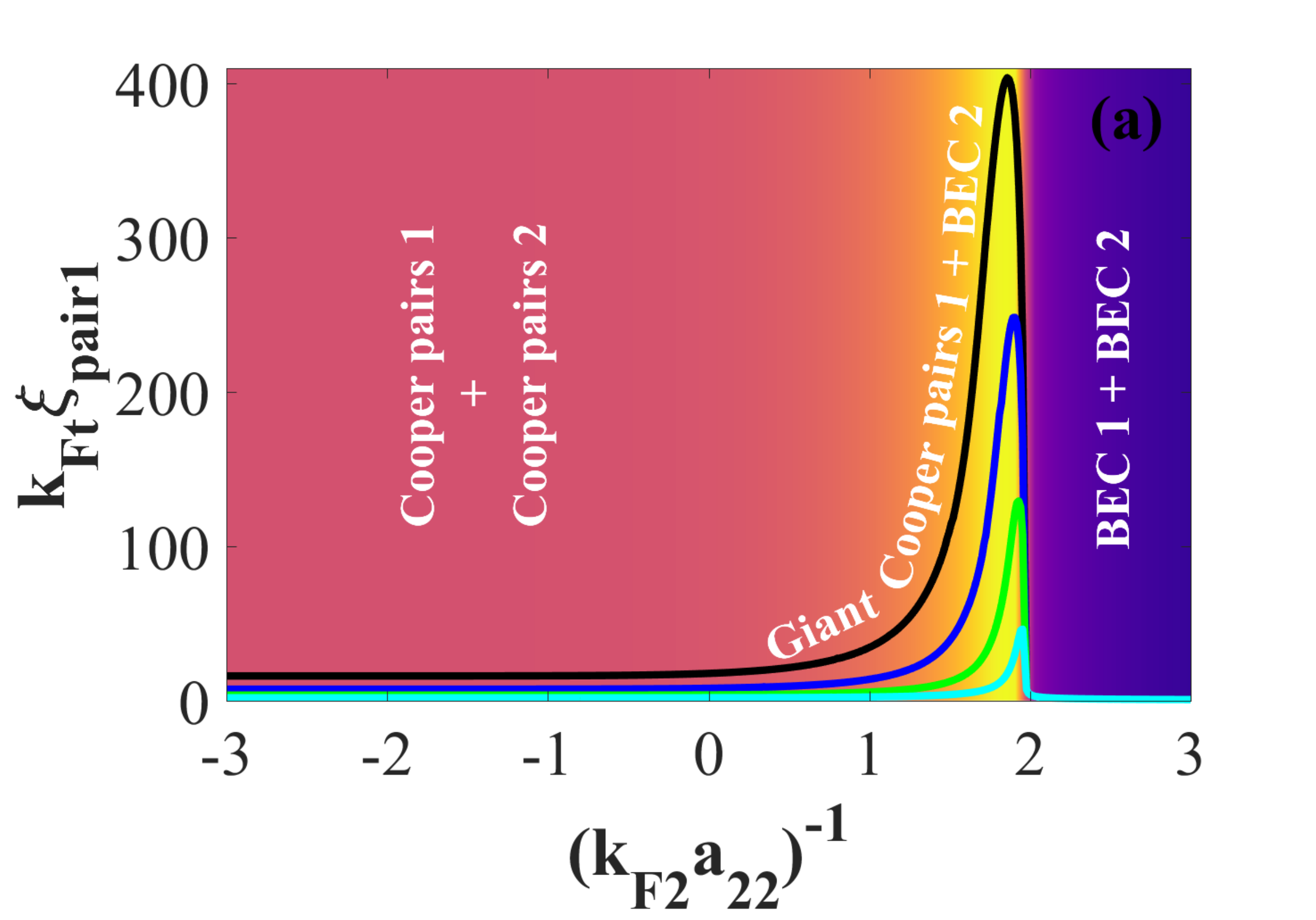}
\includegraphics[width=0.49\textwidth]{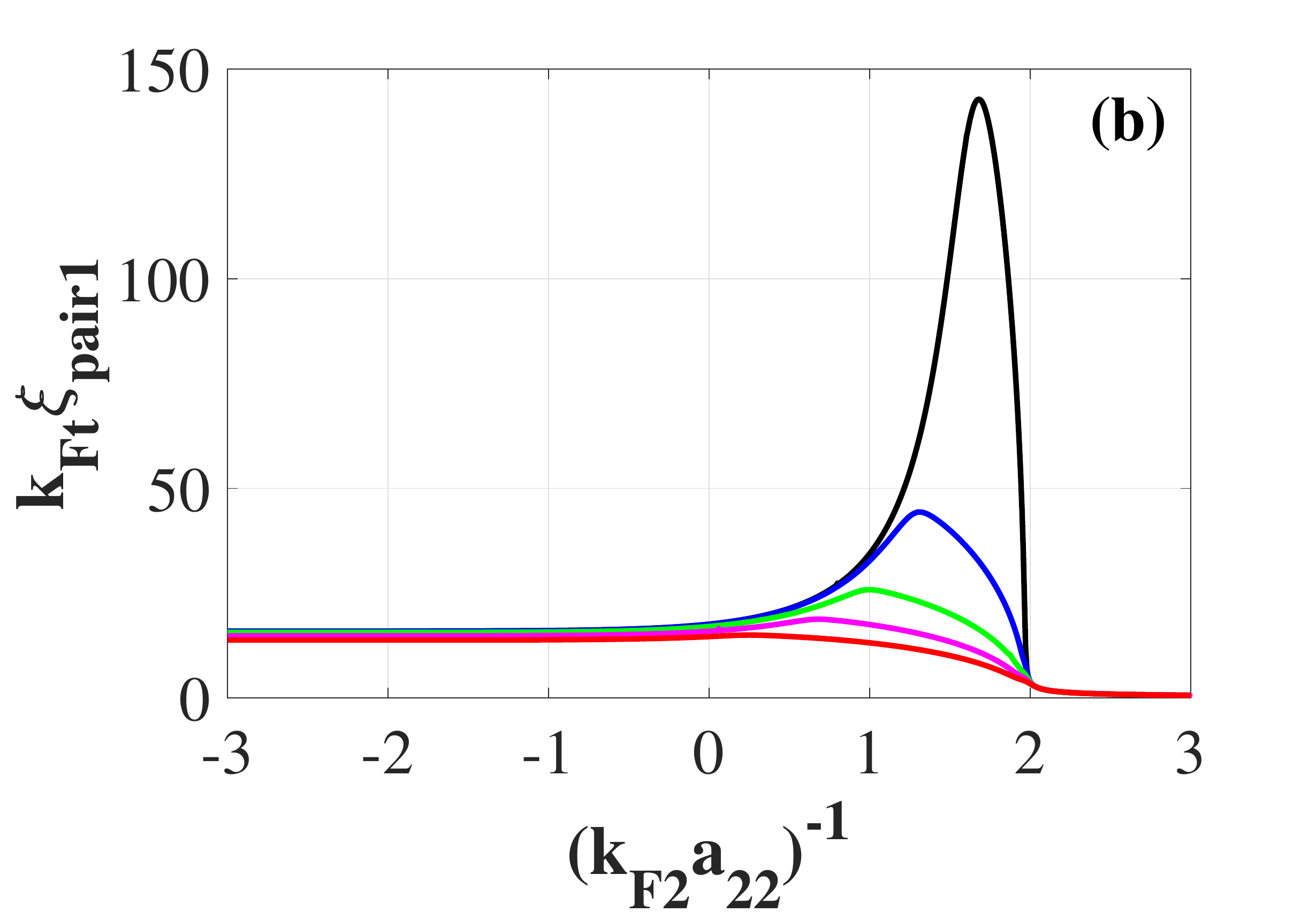}

\caption{(a) Evolution of the intrapair correlation length in the first band $\xi_{\tt{pair1}}$ of a very weakly interacting  superfluid two-band Fermi gas with $\tilde{U}_{12} = 0.001$  as a function of $(k_{\rm F2}a_{22})^{-1}$  for  $(k_{\rm F1}a_{11})^{-1} = -2$ (black line),  $(k_{\rm F1}a_{11})^{-1} = -1.5$ (blue line),  $(k_{\rm F1}a_{11})^{-1} = -1$ (green line) and $(k_{\rm F1}a_{11})^{-1} = -0.5 $ (cyan line) at $T=0$. 
(b) Temperature effect on the amplification of the intrapair correlation length  $\xi_{\tt{pair}1}$ for $\tilde{U}_{12} = 0.001$ with the fixed value $(k_{\rm F1}a_{11})^{-1} = -2$. Black line is for $T=0$ (the maximum value of the intraband length corresponds to the critical temperature of the system $T_c \approx 0.67T_{Ft}$ and to the critical temperature of the first band $T_{c1} \approx 0.0007T_{Ft}$), blue line is for $T=0.005E_{Ft}$ (maximum of $\xi_{\tt{pair}1}$ corresponds to $T_c \approx 0.523T_{Ft}$ and $T_{c1} \approx 0.0043T_{Ft}$),  green line - $T=0.01E_{Ft}$ ($T_c \approx 0.41T_{Ft}$ and $T_{c1} \approx 0.0092T_{Ft}$), cyan line - $T=0.015E_{Ft}$ ($T_c \approx 0.3T_{Ft}$ and  $T_{c1} \approx 0.014T_{Ft}$) and red line is for $T=0.02E_{Ft}$ ($T_c \approx 0.188T_{Ft}$ and $T_{c1} \approx 0.02T_{Ft}$).}
\label{fig7}
\end{figure}

Apart from the increasing of intraband and interband coupling strengths, here we show how the temperature also effects the phenomenon of giant Cooper pair formation. The slow growing of the temperature decreases the magnitude of the effect dramatically and shifts slightly the position of the peak to smaller values of $(k_{\rm F2}a_{22})^{-1}$ (Fig. 7b). The explanation of such behavior will be provided below. 

\begin{figure}
\includegraphics[width=0.49\textwidth]{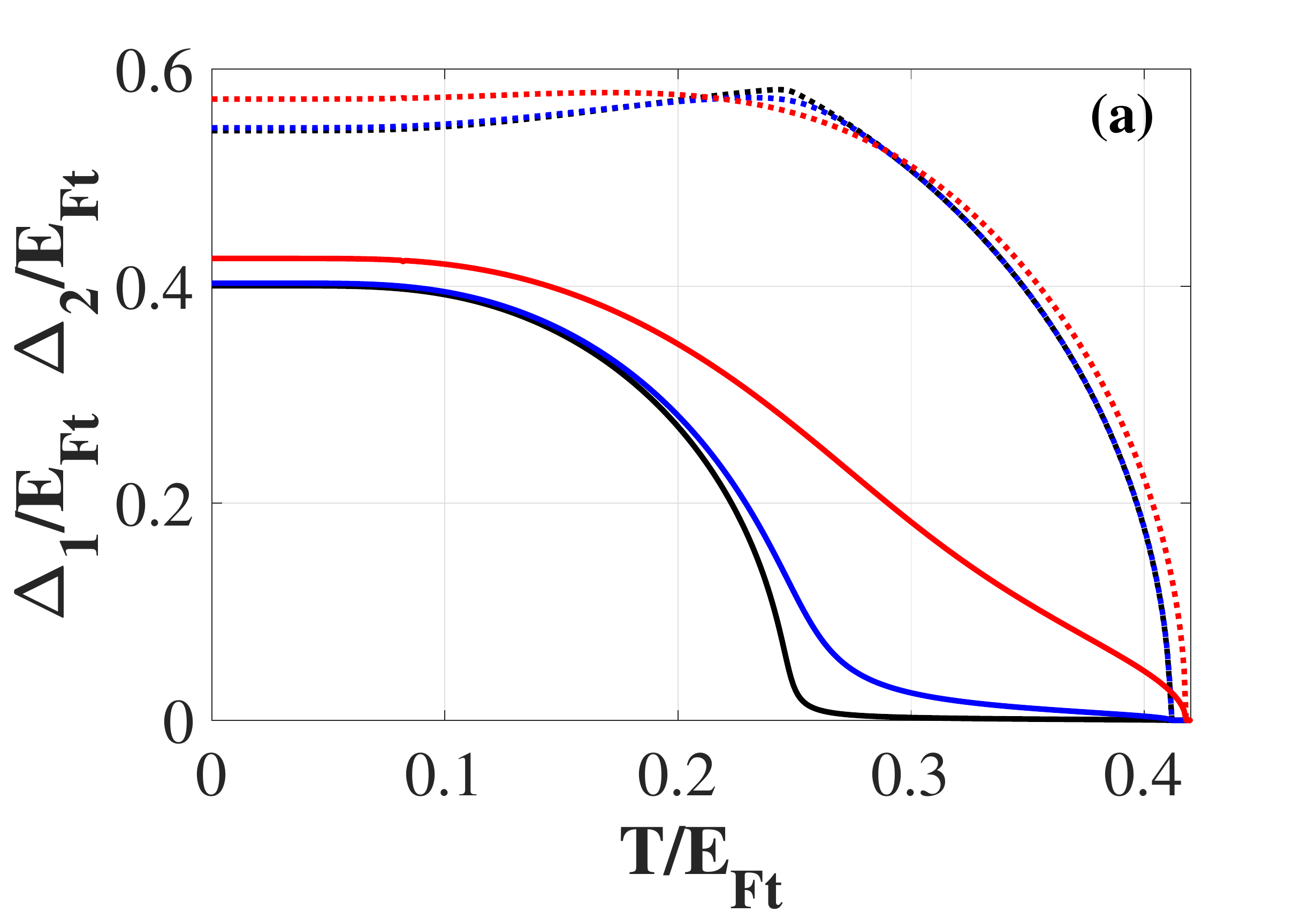}
\includegraphics[width=0.49\textwidth]{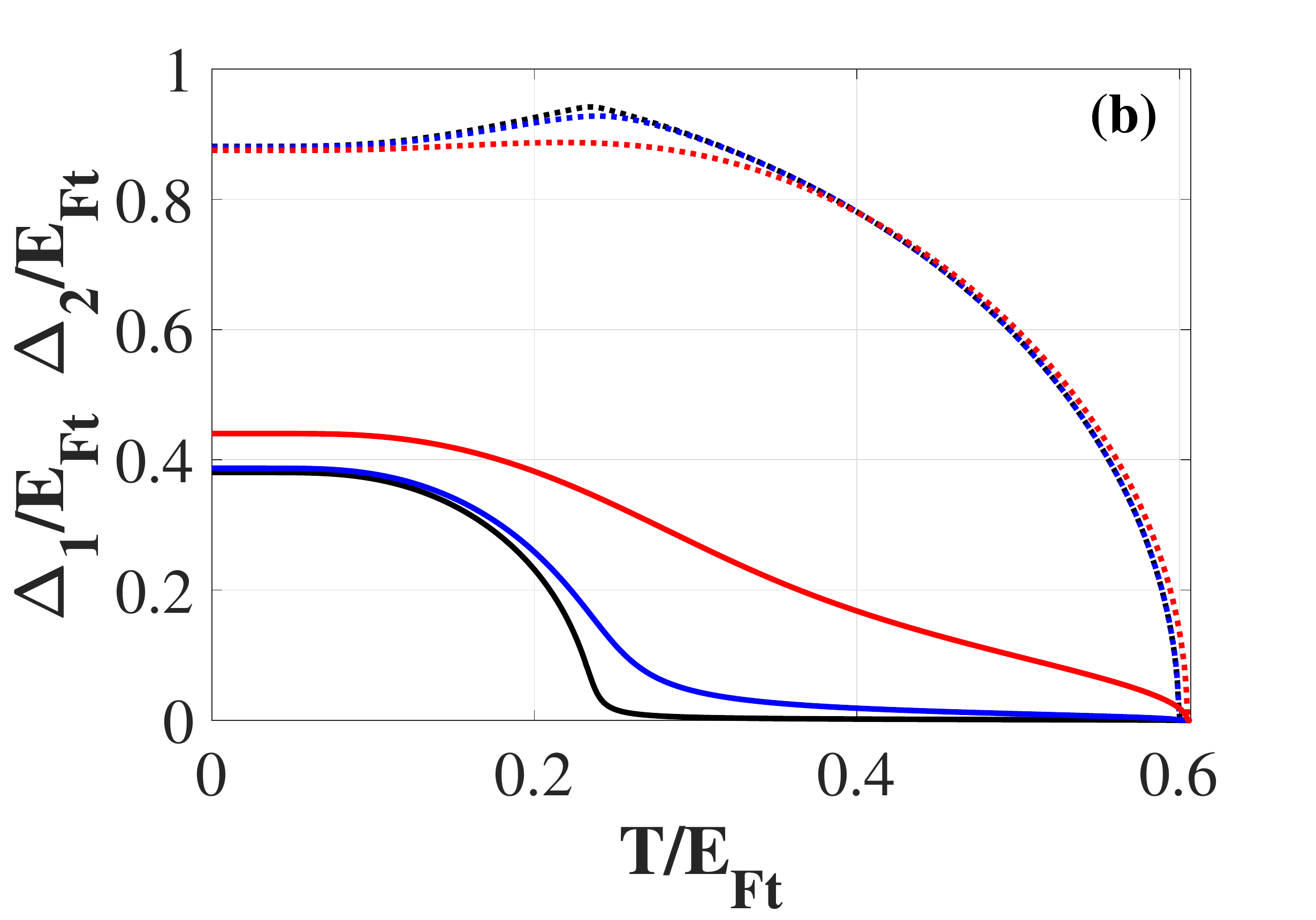}

\caption{Temperature dependence of the energy gaps $\Delta_1$ (solid line) and $\Delta_2$ (dotted line) of a two-band superfluid Fermi gas with $\tilde U_{12} = 0.001$ (black line), $\tilde U_{12} = 0.01$ (blue line) and $\tilde U_{12} = 0.1$ (red line)  for $(k_{\rm F1}a_{11})^{-1} = -0.25$, $(k_{\rm F2}a_{22})^{-1} = 1$ (a) and $(k_{\rm F1}a_{11})^{-1} = 0$, $(k_{\rm F2}a_{22})^{-1} = 1.5$ (b).}
\label{fig8}
\end{figure}

Investigations of the temperature dependences of the energy gaps show that in the case of weak interband coupling we observe anomalous behavior of  $\Delta_2$ (Fig. 8). The occurrence of the kink is directly connected with the increasing of the intrapair correlation length in the first band. Moreover, temperature dependences of the chemical potential (not shown) also have a kink for the same values of $T$ and hence indicate the first order phase transition in the system. 
Numerical analysis shows that the effect is more pronounced when the values of energy gaps become comparable. At the same time we recall that the strong enhancement of $\xi_{\tt{pair}1}$ is realized in the BEC limit of the second band at $T=0$  and for vanishing $U_{12}$ (Fig. 7a). The increasing of the temperature, the strong interband interaction and the strong intraband coupling in the first band lead to the suppression of the intrapair correlation length (Figs. 6 and  7b). Thus, comparability of energy gaps with the conservation of the intrapair correlation length amplification effect can be achieved, when the second band is in the strong-coupling regime (for values of  $(k_{\rm F2}a_{22})^{-1}$, where we have the strong enhancement of $\xi_{\tt{pair}1}$) and when the first band is near the unitarity point $(k_{\rm F1}a_{11})^{-1}  \approx 0$ (this yields ${\Delta _1} \cong {\Delta _2}$ ). 
Other conditions eliminate the non-monotonic dependence of ${\Delta _2 (T)}$, leading to the conventional BCS-like behavior of the energy gap in the second band, or sufficiently decrease the magnitude of the peak. In turn with the increase of the temperature the suppression of the peak in $\xi_{\tt{pair}1}$ becomes important when the temperature is in the vicinity of the critical temperature of the first cold band $T_{c1}$, which is the very small for the case of parameters here considered (see the legend in the Fig. 7b).

Based on this, we can claim that the temperature decreasing rate of the intrapair correlation length in the first band is determined by the width of the temperature interval, where the non-BCS behavior of the second gap is realized (the same, where the first energy gap is not strongly suppressed). Since Fig. 7b corresponds to the system with  $(k_{\rm F1}a_{11})^{-1}=-2$, i.e. the BCS regime for the first band, we have the small temperature interval of non-BCS dependence of ${\Delta _2}$ and as a consequence the rapid temperature suppression of $\xi_{\tt{pair}1}$. 

It is important to note that similar behavior of the intrapair correlation lengths as a function of temperature was revealed in a two-band superconductor with very weak interband interaction \cite{Komendova}.  In the absence of coupling between two superconducting condensates below the critical temperature, a hidden critical point appears at the critical temperature of the weaker band that corresponds to the divergence of the intrapair correlation length. In the case of weak interband interaction the intrapair correlation length of the weaker band exhibits a deviation from the conventional monotonic increase with temperature and leads to a pronounced peak close to the hidden critical point. In our calculations interband coupling also governs the effect but as opposite to a two-band superconductor strong enhancement of the intrapair correlation length in one of the bands occurs in the strong-coupling limit, where the formation of giant Cooper pairs is not expected. Moreover as it was shown above this phenomenon for very weak interband interaction can be observed even for at finite temperatures. We emphasize that the hidden critical-like behavior via the temperature dependence of the energy gap ${\Delta _1}$ in the cold band in the strong-coupling limit, whereas monotonic behavior of the energy gap was reported in the weak-coupling limit in Ref. \onlinecite{Komendova}.

We suggest that the experimental detection of giant Cooper pairs in the strong-coupling regime and verification of our prediction can be done through direct imaging of vortex cores in two-component fermionic condensates or in iron-based superconductors with electronic-like concentric Fermi surfaces. Another possibility to verify our predictions is the precise measurement of the temperature dependence of the energy gaps in two-band superfluid systems.

\section{Conclusions}

We have investigated characteristics and have found novel unique properties of the BCS-BEC crossover in a two-band superfluid Fermi system in the presence of energy shift between bands within a mean-field theory in a configuration of different pairing strengths in the two bands. We  have demonstrated the richness of the BCS-BEC crossover in such a two-band system as compared to the single-band counterpart. We have found that for vanishing interband interaction at low temperatures and in the strong-coupling regime, a two-band superfluid Fermi gas evolves to a single-band system with the full suppression of the energy gap in the first cold band, together with the full redistribution of particles. As a result, a giant enhancement of the intrapair correlation length of Cooper pairs in the first band occurs. In the case of finite coupling between the two condensates of the two bands, we have shown a non-monotonic behavior of the first energy gap with a hump, the position of which is determined by the strength of the interband interaction in the second hot band. For weak interband coupling we have found a significant amplification of the intrapair correlation length of the first band in the BEC regime for the second band at zero and finite temperatures that indicates the coexistence of giant Cooper pairs and bosonic condensate in a two-band superfluid system. We have revealed that such an effect can produce an unusual non-monotonic temperature dependence of the second energy gap with the presence of a maximum for nonzero temperatures. Our predictions can be verified via STM investigations of vortex cores and temperature behavior of energy gaps in  two-component atomic condensates and in some iron-based superconductors having electron-like or hole-like concentric bands with low filling, with weak interband interaction between the bands. 

\acknowledgments

This work was supported by the Italian MIUR through the PRIN 2015 program (Contract No. 2015C5SEJJ001). H. T. was supported by Grant-in-Aid for JSPS fellows (Grant No.  17J03975). We thank Alexei Vagov and Milorad V. Milo\v{s}evi\'c  for discussions.
\appendix
\section{Critical temperature, chemical potential and particle densities}
To simplify Eqs. (3)-(7) in the main paper for the numerical solution and analysis we will measure $k$, energy gaps $\Delta _i$ and the chemical potential $\mu$ in units of the total Fermi momentum $k_{Ft}$, the total Fermi energy $E_{Ft}$ and the temperature $T/E_{Ft}$ 

\begin{widetext}
\begin{eqnarray}
\label{eqS1}
{{\tilde \Delta }_1}\int\limits_0^{{k_0}/{k_{Ft}}} {{x^2}\frac{{\tanh \frac{{\sqrt {{{\left( {{x^2} - \tilde \mu } \right)}^2} + \tilde \Delta _1^2} }}{{2t}}}}{{\sqrt {{{\left( {{x^2} - \tilde \mu } \right)}^2} + \tilde \Delta _1^2} }}} dx = \frac{{{A_{11}}}}{W}{{\tilde \Delta }_1} - \frac{3}{4}\frac{{{{\tilde U}_{12}}}}{W}{\left( {\frac{{{k_{Ft}}}}{{{k_0}}}} \right)^2}{{\tilde \Delta }_2},
\end{eqnarray}
\begin{eqnarray}
\label{eqS2}
{\tilde \Delta _2}\int\limits_0^{{k_0}/{k_{Ft}}} {{x^2}\frac{{\tanh \frac{{\sqrt {{{\left( {{x^2} - \tilde \mu  + \frac{{{E_g}}}{{{E_{Ft}}}}} \right)}^2} + \tilde \Delta _2^2} }}{{2t}}}}{{\sqrt {{{\left( {{x^2} - \tilde \mu  + \frac{{{E_g}}}{{{E_{Ft}}}}} \right)}^2} + \tilde \Delta _2^2} }}} dx = \frac{{{A_{11}}}}{W}{\tilde \Delta _2} - \frac{3}{4}\frac{{{{\tilde U}_{21}}}}{W}{\left( {\frac{{{k_{Ft}}}}{{{k_0}}}} \right)^2}{\tilde \Delta _1},
\end{eqnarray}
\begin{eqnarray}
\label{eqS3}
\frac{2}{3} = \int\limits_0^{{k_0}/{k_{Ft}}} {{x^2}} \left( {1 - \frac{{{x^2} - \tilde \mu }}{{\sqrt {{{\left( {{x^2} - \tilde \mu } \right)}^2} + \tilde \Delta _1^2} }}\tanh \frac{{\sqrt {{{\left( {{x^2} - \tilde \mu } \right)}^2} + \tilde \Delta _1^2} }}{{2t}}} \right)dx + \nonumber \\
\int\limits_0^{{k_0}/{k_{Ft}}} {{x^2}\left( {1 - \frac{{{x^2} - \tilde \mu  + \frac{{{E_g}}}{{{E_{Ft}}}}}}{{\sqrt {{{\left( {{x^2} - \tilde \mu  + \frac{{{E_g}}}{{{E_{Ft}}}}} \right)}^2} + \tilde \Delta _2^2} }}\tanh \frac{{\sqrt {{{\left( {{x^2} - \tilde \mu  + \frac{{{E_g}}}{{{E_{Ft}}}}} \right)}^2} + \tilde \Delta _2^2} }}{{2t}}} \right)dx},
\end{eqnarray}
\end{widetext}
where $x = k/k_{Ft}$, $\tilde \Delta _i = \Delta _i/E_{Ft}$, $\tilde \mu = \mu/E_{Ft}$ and $t=T/E_{Ft}$. Also we introduce a notation $W = {A_{11}}{A_{22}} - \frac{9}{{16}}{{\tilde U}_{12}}{{\tilde U}_{21}}{\left( {\frac{{{k_{Ft}}}}{{{k_0}}}} \right)^4}$. For the sake of simplicity we will omit the tilde symbol in Eqs. (\ref{eqS1})-(\ref{eqS3}) for the dimensionless energy gaps and the chemical potential.

In the vicinity of the critical temperature we can linearize the system of Eqs. (\ref{eqS1})-(\ref{eqS3}) and from the condition of the solvability we obtain the equation for the critical temperature and the particle densities
\begin{widetext}
\begin{eqnarray}
\label{eqS4}
\left( {\int\limits_0^{{k_0}/{k_{Ft}}} {\frac{{{x^2}\tanh \frac{{{x^2} - \mu }}{{2t}}}}{{{x^2} - \mu }}} dx - \frac{{{A_{22}}}}{W}} \right) \times \left( {\int\limits_0^{{k_0}/{k_{Ft}}} {\frac{{{x^2}\tanh \frac{{{x^2} - \mu  + \frac{{{E_g}}}{{{E_{Ft}}}}}}{{2t}}}}{{{x^2} - \mu  + \frac{{{E_g}}}{{{E_{Ft}}}}}}} dx - \frac{{{A_{11}}}}{W}} \right) - \frac{9}{{16}}\frac{{{{\tilde U}_{12}}{{\tilde U}_{21}}}}{{{W^2}}}{\left( {\frac{{{k_{Ft}}}}{{{k_0}}}} \right)^4} = 0,
\end{eqnarray}
\begin{eqnarray}
\label{eqS5}
\frac{2}{3} = \int\limits_0^{{k_0}/{k_{Ft}}} {{x^2}} \left( {1 - \tanh \frac{{{x^2} - \mu }}{{2t}}} \right)dx + \int\limits_0^{{k_0}/{k_{Ft}}} {{x^2}\left( {1 - \tanh \frac{{{x^2} - \mu  + \frac{{{E_g}}}{{{E_{Ft}}}}}}{{2t}}} \right)dx}.
\end{eqnarray}
\end{widetext}
Based on asymptotic expansions for $\frac{{{k_0}}}{{{k_{Ft}}}} \gg 1$ 
\begin{equation}
\label{eqS6}
\frac{{{k_0}}}{{{k_{Ft}}}} - \frac{{{A_{11}}}}{W} \approx \frac{\pi }{{2{k_{F1}}{a_{11}}}}\frac{{{k_{F1}}}}{{{k_{Ft}}}}
\end{equation}
and
\begin{equation}
\label{eqS7}
\frac{{{k_0}}}{{{k_{Ft}}}} - \frac{{{A_{22}}}}{W} \approx \frac{\pi }{{2{k_{F2}}{a_{22}}}}\frac{{{k_{F2}}}}{{{k_{Ft}}}}
\end{equation}
we rewrite the Eqs. (\ref{eqS4}) and (\ref{eqS5}) extending the limit of the integration up to infinity and thereby eliminating the cut-off momentum dependence for the determination of the critical temperature
\begin{widetext}
\begin{eqnarray}
\label{eqS8}
\left( {\int\limits_0^{ + \infty } {\left( {1 - \frac{{{x^2}\tanh \frac{{{x^2} - \mu }}{{2t}}}}{{{x^2} - \mu }}} \right)} dx - \frac{\pi }{{2{k_{F1}}{a_{11}}}}\frac{{{k_{F1}}}}{{{k_{Ft}}}}} \right) \times & \nonumber \\ \left( {\int\limits_0^{ + \infty } {\left( {1 - \frac{{{x^2}\tanh \frac{{{x^2} - \mu  + \frac{{{E_g}}}{{{E_{Ft}}}}}}{{2t}}}}{{{x^2} - \mu  + \frac{{{E_g}}}{{{E_{Ft}}}}}}} \right)} dx - \frac{\pi }{{2{k_{F2}}{a_{22}}}}\frac{{{k_{F2}}}}{{{k_{Ft}}}}} \right) - \frac{9}{{16}}{\tilde U_{12}}{\tilde U_{21}} = 0,
\end{eqnarray}
\begin{eqnarray}
\label{eqS8_5}
- {t^{\frac{3}{2}}}\Gamma \left( {\frac{3}{2}} \right){\rm{L}}{{\rm{i}}_{\frac{3}{2}}}\left( { - {e^{\frac{\mu }{t}}}} \right) - {t^{\frac{3}{2}}}\Gamma \left( {\frac{3}{2}} \right){\rm{L}}{{\rm{i}}_{\frac{3}{2}}}\left( { - {e^{\frac{{\mu  - \frac{{{E_g}}}{{{E_{Ft}}}}}}{t}}}} \right) = \frac{2}{3},
\end{eqnarray}
\end{widetext}
where the intergration of Eq. (\ref{eqS5}) was performed in terms of the polylogarithm function ${\rm{L}}{{\rm{i}}_s}\left( z \right)$ and the gamma function $\Gamma(z)$. Taking into account that 
\begin{equation}
\label{eqS8_6}
\mathop {\lim }\limits_{t \to 0} \left( { - {t^{\frac{3}{2}}}{\rm{L}}{{\rm{i}}_{\frac{3}{2}}}\left( { - {e^{\frac{\mu }{t}}}} \right)} \right) = \frac{4}{{3\sqrt \pi  }}{\mu ^{\frac{3}{2}}},
\end{equation}
we simplify Eq.  (\ref{eqS8_5}) in the BCS limit when $t \ll 1$
\begin{equation}
\label{eqS8_7}
{\mu_{s} ^{\frac{3}{2}}} + {\left( {\mu_{s}  - \frac{{{E_g}}}{{{E_{Ft}}}}} \right)^{\frac{3}{2}}} = 1,
\end{equation}
which has an approximated solution
\begin{equation}
\label{eqS8_8}
\mu_{s}  \approx \frac{1}{3}\frac{{3\sqrt {1 - \frac{{{E_g}}}{{{E_{Ft}}}}}  - 2{{\left( {1 - \frac{{{E_g}}}{{{E_{Ft}}}}} \right)}^{\frac{3}{2}}} + 3}}{{1 + \sqrt {1 - \frac{{{E_g}}}{{{E_{Ft}}}}} }}.
\end{equation}
For the zero energy shift $E_g = 0$ Eq. (\ref{eqS8}) is reduced to equation
\begin{eqnarray}
\label{eqS9}
({\ln \frac{{{2^{ \frac{2}{3}}}\pi {e^2}t}}{{8e^{\gamma} }} - \frac{\pi }{{2{k_{F1}}{a_{11}}}}\frac{{{k_{F1}}}}{{{k_{Ft}}}}}) \times   \nonumber \\ ( {\ln \frac{{{2^{  \frac{2}{3}}}\pi {e^2}t}}{{8e^{\gamma} }} - \frac{\pi }{{2{k_{F2}}{a_{22}}}} \frac{{{k_{F2}}}}{{{k_{Ft}}}}}) - \frac{9}{{16}}{{\tilde U}_{12}}{{\tilde U}_{21}} = 0,
\end{eqnarray}
where $\gamma = 0.577...$ is the Euler-Mascheroni constant and where the simplified equation for the weak-coupling limit Eq.  (\ref{eqS8_7}) gives the exact value of the chemical potential $\mu = {2^{ - \frac{2}{3}}}$. Formally  Eq. (\ref{eqS9}) has the same form as a equation for the determination of the critical temperature of a clean two-band superconductor \cite{Suhl, Moskalenko, Gurevich} with the corresponding solution
\begin{widetext}
\begin{equation}
\label{eqS9-5}
\frac{{{T_c}}}{{{E_{Ft}}}} = \frac{{8e^{\gamma}}}{{ {2^{ \frac{2}{3}}}\pi {e^2}}}\exp \left( {\frac{1}{4}\left( {\frac{\pi }{{{k_{F1}}{a_{11}}}}\frac{{{k_{F1}}}}{{{k_{Ft}}}} + \frac{\pi }{{{k_{F2}}{a_{22}}}}\frac{{{k_{F2}}}}{{{k_{Ft}}}} + \sqrt {{{\left( {\frac{\pi }{{{k_{F1}}{a_{11}}}}\frac{{{k_{F1}}}}{{{k_{Ft}}}} - \frac{\pi }{{{k_{F2}}{a_{22}}}}\frac{{{k_{F2}}}}{{{k_{Ft}}}}} \right)}^2} + 9{{\tilde U}_{12}}{{\tilde U}_{21}}} } \right)} \right).
\end{equation}
\end{widetext}
For the opposite case when ${E_g} \gg 1$, i.e. with the large separation between energy bands we can neglect a hyperbolic tangent function in the second bracket expression of Eq. (\ref{eqS8}). This yields the equation and the solution
\begin{flalign}
\label{eqS10}
({\ln \frac{{\pi {e^2}t}}{{8\mu_{s}e^{\gamma} }} - \frac{\pi }{{2{k_{F1}}{a_{11}}}}\frac{{{k_{F1}}}}{{{k_{Ft}}}}} ) \times & \nonumber \\  ( {\frac{\pi }{2}\sqrt {\frac{{{E_g}}}{{{E_{Ft}}}} - 1}  - \frac{\pi }{{2{k_{F2}}{a_{22}}}}\frac{{{k_{F2}}}}{{{k_{Ft}}}}}) - \frac{9}{{16}}{{\tilde U}_{12}}{{\tilde U}_{21}} = 0,
\end{flalign}
\begin{widetext}
\begin{equation}
\label{eqS10-5}
\frac{{{T_c}}}{{{E_{Ft}}}} = \frac{{8\mu_{s}e^{\gamma} }}{{\pi {e^2}}} \exp \left( {\frac{\pi }{{2{k_{F1}}{a_{11}}}}\frac{{{k_{F1}}}}{{{k_{Ft}}}} + \frac{9}{{16}}\frac{{{{\tilde U}_{12}}{{\tilde U}_{21}}}}{{\frac{\pi }{2}\sqrt {\frac{{{E_g}}}{{{E_{Ft}}}} - 1}   - \frac{\pi }{{2{k_{F2}}{a_{22}}}}\frac{{{k_{F2}}}}{{{k_{Ft}}}}}}} \right).
\end{equation}
\end{widetext}
We can see that in the case of vanishing interaction between bands for the large shift between bands the critical temperature of the system is independent on the $1/k_{F2}a_{22}$ and in the BCS regime is determined by the first band only with the reduced value of the chemical potential $\mu_{s}$. 

Now we analyze numerically the general behavior of the critical temperature and the chemical potential of a two-band superfluid Fermi gas in the vicinity of $T_c$. As we can see with the increasing of the interband coupling strength we observe an increase of the critical temperature with the corresponding decrease of the chemical potential (Fig. 9). 
\begin{figure*}
\includegraphics[width=0.49\textwidth]{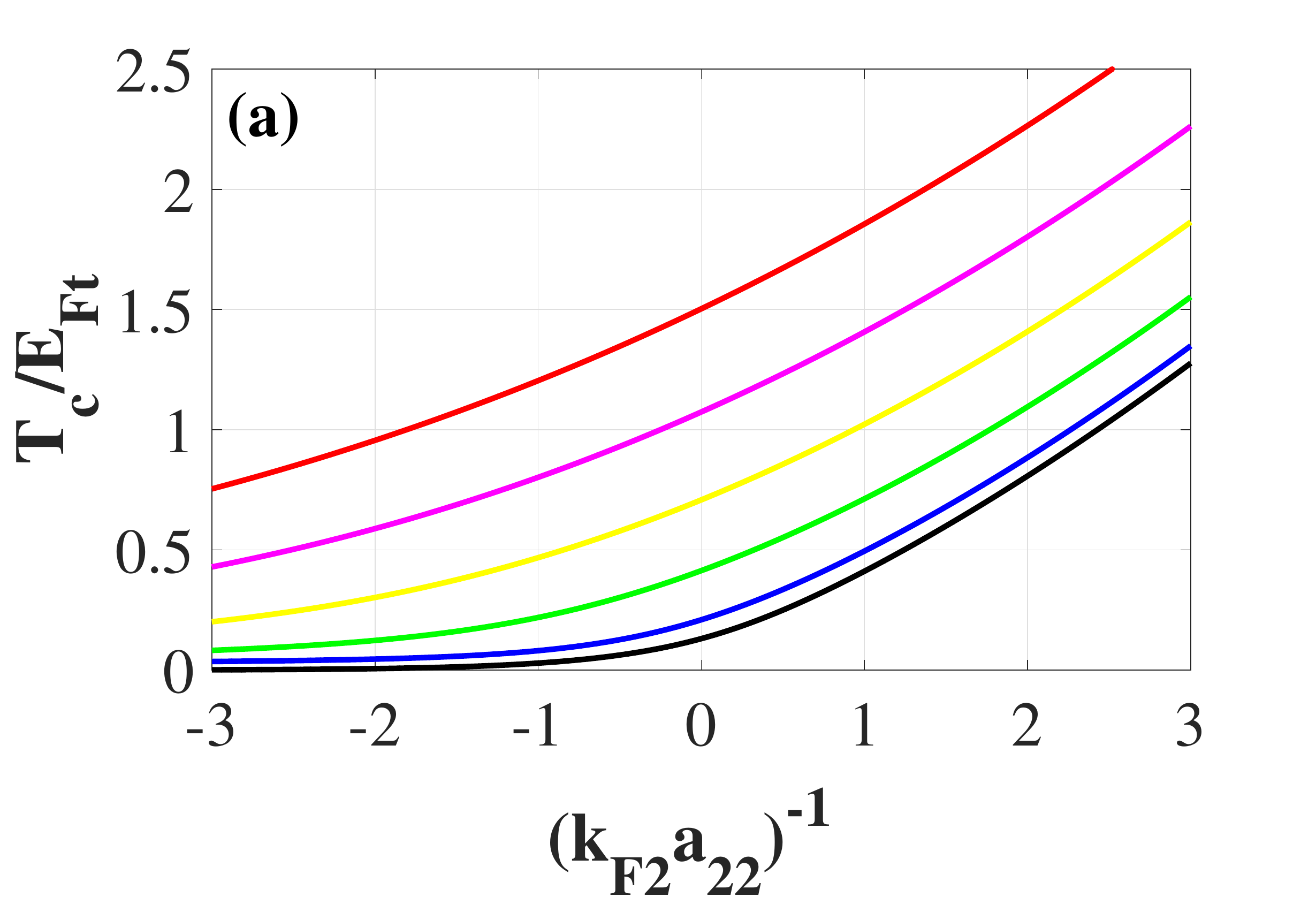}
\includegraphics[width=0.49\textwidth]{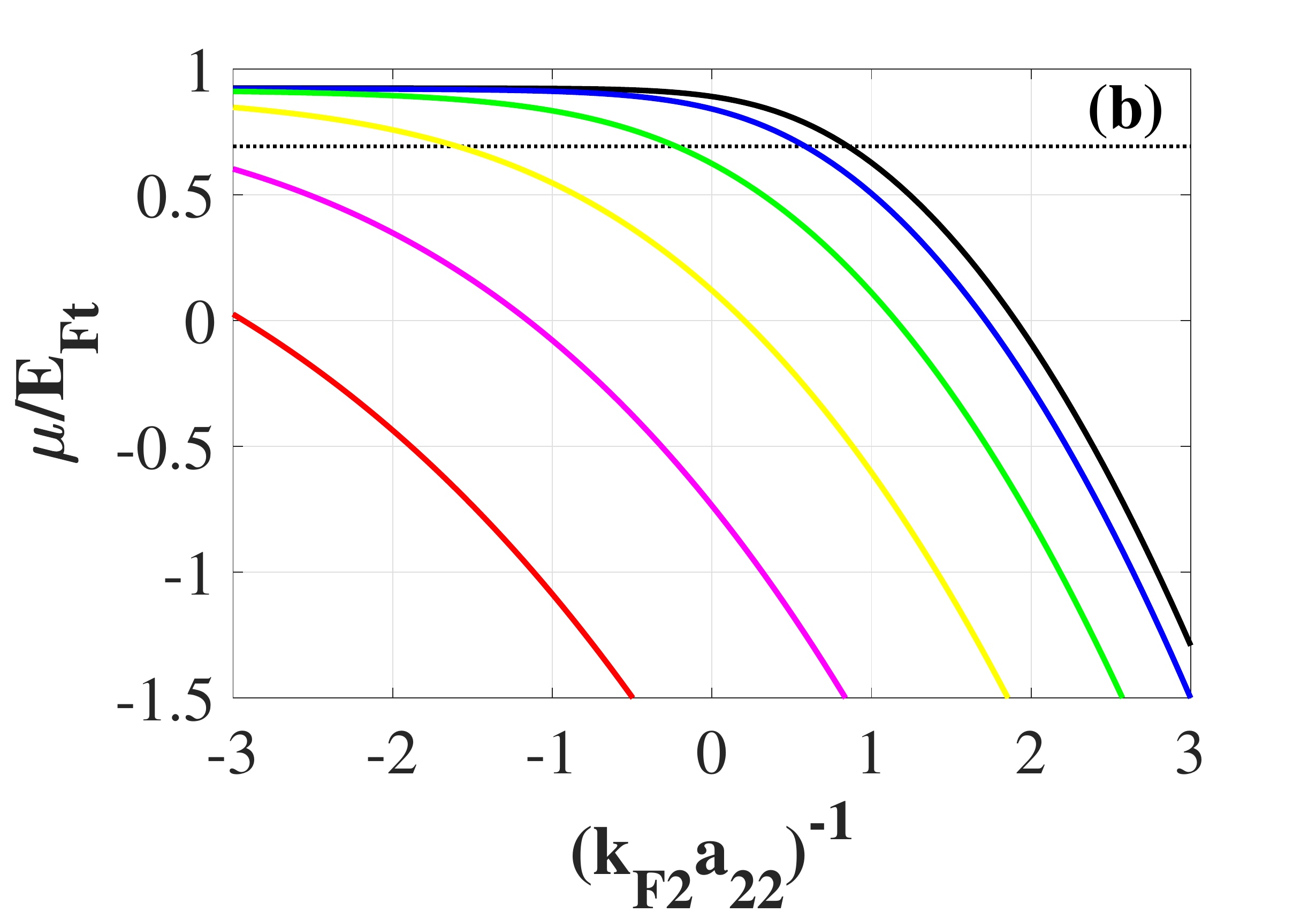}
\caption{(a) The critical temperature $T_{\rm c}$ and (b) the chemical potential $\mu$ (at $T=T_{\rm c}$) vs $(k_{\rm F2}a_{22})^{-1}$ for different interband couplings  $\tilde{U}_{12} = 0$ (black line), $\tilde{U}_{12} = 1$ (blue line), $\tilde{U}_{12} = 2$ (green line), $\tilde{U}_{12} = 3$ (yellow line), $\tilde{U}_{12} = 4$ (magenta line),  $\tilde{U}_{12} = 5$ (red line) with the fixed value of the scattering length in the first band $(k_{\rm F1}a_{11})^{-1} = -2$. Dotted black line is the energy shift $E_g$ between bands in units of $E_{Ft}$.}
\label{fig9}
\end{figure*}

\begin{figure}
\includegraphics[width=0.99\columnwidth]{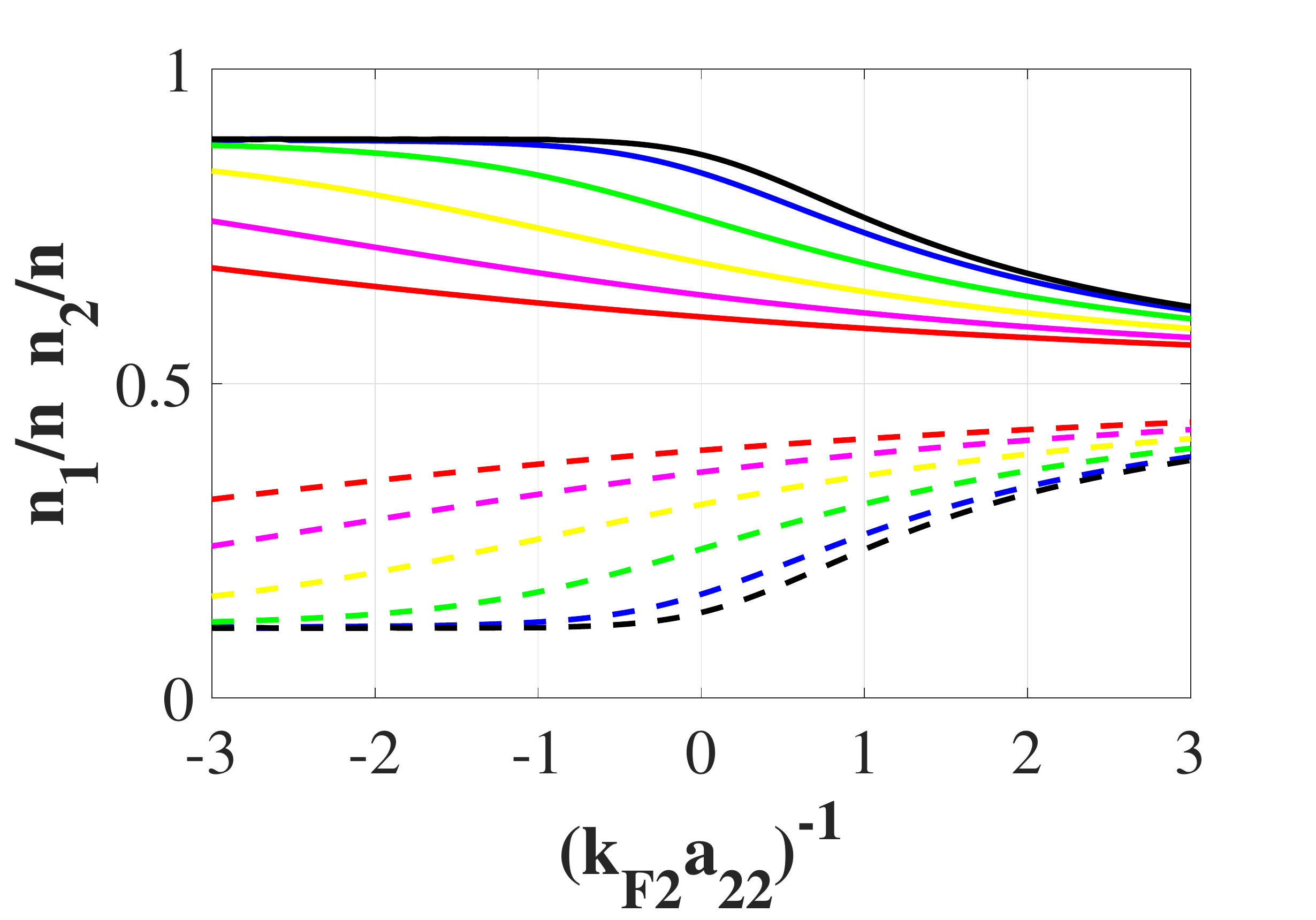}
\caption{Particle densities in the first (solid lines) and in the second (dashed lines) bands for different interband couplings  $\tilde{U}_{12} = 0$ (black line), $\tilde{U}_{12} = 1$ (blue line), $\tilde{U}_{12} = 2$ (green line), $\tilde{U}_{12} = 3$ (yellow line), $\tilde{U}_{12} = 4$ (magenta line),  $\tilde{U}_{12} = 5$ (red line) with the fixed value of the scattering length in the first band $(k_{\rm F1}a_{11})^{-1} = -2$.}
\label{fig10}
\end{figure}
Fig. 10 shows the distribution of particle densities between bands with the increasing of the interband coupling strength near the critical temperature. For the strong interband interaction dependences $n_1$ and $n_2$ exhibit a tendency to equalizing of particle densities in each bands towards the BEC limit of the second band. Comparing with the Fig. 5 in the paper \cite{Tajima} one can see that pairing  fluctuations associated with both interband and intraband couplings reduced significantly the effects of the particle interband distribution in the vicinity of $T_c$. 

Also it should be noted that the effect of the interband interaction is the most distinct for low temperatures, where even small changes of $U_{12}$ gives rise a perceptible effect for the energy gaps and the chemical potential (see fig. 2 in the main paper).

\section{Comparison of approaches: Fixed intraband couplings ratio vs fixed intraband coupling in one of the band}

In this section we compare two different approaches for the description of the BCS-BEC crossover characteristics based on a mean-field theory and their effect on the results. We start from the simple case: two-band superfluid Fermi gas with vanishing interaction between bands at zero temperature. 
At the beginning we consider the approach with the fixed ratio of intraband couplings $\rho = \tilde U_{22}/\tilde U_{11}$.  This definition gives the relation between scattering lengths in each band
\begin{equation}
\label{eqS11}
\frac{\pi }{{2{k_{F1}}{a_{11}}}}\frac{{{k_{F1}}}}{{{k_{Ft}}}} =  - \frac{{{k_0}}}{{{k_{Ft}}}}\left( {\rho  - 1} \right) + \frac{\pi }{{2{k_{F2}}{a_{22}}}}\frac{{{k_{F2}}}}{{{k_{Ft}}}}\rho 
\end{equation}
Substituting of Eq. (\ref{eqS11}) to Eqs. (3)-(5) in the main paper for the energy gaps and the particle densities  and taking into account dimensionless notations we have again
\begin{widetext}
\begin{eqnarray}
\label{eqS12}
{{\Delta }_1}\int\limits_0^{{k_0}/{k_{Ft}}} {\frac{{{x^2}}}{{\sqrt {{{\left( {{x^2} - \mu } \right)}^2} +  \Delta _1^2} }}} dx = \frac{{\frac{1}{{\frac{{{k_0}}}{{{k_{Ft}}}} - \frac{\pi }{{2{k_{F2}}{a_{22}}}}\frac{{{k_{F2}}}}{{{k_{Ft}}}}}}{{ \Delta }_1} - \frac{3}{4}{{\tilde U}_{12}}{{\left( {\frac{{{k_{Ft}}}}{{{k_0}}}} \right)}^2}{{ \Delta }_2}}}{{\frac{1}{{\frac{{{k_0}}}{{{k_{Ft}}}} + \frac{{{k_0}}}{{{k_{Ft}}}}\left( {\rho  - 1} \right) - \frac{\pi }{{2{k_{F2}}{a_{22}}}}\frac{{{k_{F2}}}}{{{k_{Ft}}}}\rho }} \cdot \frac{1}{{\frac{{{k_0}}}{{{k_{Ft}}}} - \frac{\pi }{{2{k_{F2}}{a_{22}}}}\frac{{{k_{F2}}}}{{{k_{Ft}}}}}} - {{\tilde U}_{12}}{{\tilde U}_{21}}{{\left( {\frac{{{k_{Ft}}}}{{{k_0}}}} \right)}^4}}},
\end{eqnarray}
\begin{eqnarray}
\label{eqS13}
{{ \Delta }_2}\int\limits_0^{{k_0}/{k_{Ft}}} {\frac{{{x^2}}}{{\sqrt {{{\left( {{x^2} - \mu  + \frac{{{E_g}}}{{{E_{Ft}}}}} \right)}^2} +  \Delta _2^2} }}} dx  = \frac{{\frac{1}{{\frac{{{k_0}}}{{{k_{Ft}}}} + \frac{{{k_0}}}{{{k_{Ft}}}}\left( {\rho  - 1} \right) - \frac{\pi }{{2{k_{F2}}{a_{22}}}}\frac{{{k_{F2}}}}{{{k_{Ft}}}}\rho }}{{ \Delta }_2} - \frac{3}{4}{{\tilde U}_{21}}{{\left( {\frac{{{k_{Ft}}}}{{{k_0}}}} \right)}^2}{{ \Delta }_1}}}{{\frac{1}{{\frac{{{k_0}}}{{{k_{Ft}}}} + \frac{{{k_0}}}{{{k_{Ft}}}}\left( {\rho  - 1} \right) - \frac{\pi }{{2{k_{F2}}{a_{22}}}}\frac{{{k_{F2}}}}{{{k_{Ft}}}}\rho }} \cdot \frac{1}{{\frac{{{k_0}}}{{{k_{Ft}}}} - \frac{\pi }{{2{k_{F2}}{a_{22}}}}\frac{{{k_{F2}}}}{{{k_{Ft}}}}}} - \frac{9}{{16}}{{\tilde U}_{12}}{{\tilde U}_{21}}{{\left( {\frac{{{k_{Ft}}}}{{{k_0}}}} \right)}^4}}}
\end{eqnarray}
\begin{eqnarray}
\label{eqS14}
\frac{2}{3} = \int\limits_0^{{k_0}/{k_{Ft}}} {{x^2}} \left( {1 - \tanh \frac{{{x^2} - \mu }}{{2t}}} \right)dx + \int\limits_0^{{k_0}/{k_{Ft}}} {{x^2}\left( {1 - \tanh \frac{{{x^2} - \mu  + \frac{{{E_g}}}{{{E_{Ft}}}}}}{{2t}}} \right)dx}.
\end{eqnarray}
\end{widetext}
In comparison with the Eqs.  (\ref{eqS1}) and (\ref{eqS2}) the first two equations of the system are integrated analytically and after long but straightforward calculation is expressed via elliptic integrals
\begin{widetext}
\label{eqS15}
\begin{flalign}
\int\limits_0^{{k_0}/{k_{Ft}}} {\frac{{{x^2}}}{{\sqrt {{{\left( {{x^2} - \mu } \right)}^2} + \Delta _1^2} }}} dx  = \sqrt {i{{\Delta }_1} + \mu }  \times  \nonumber \\
\left( {F\left( {\frac{{{k_0}}}{{{k_{Ft}}}}\sqrt {\frac{{i{{\Delta }_1} + \mu }}{{\Delta _1^2 + {\mu ^2}}}} ,\sqrt {\frac{{i\mu  + {{\Delta }_1}}}{{i\mu  - {{ \Delta }_1}}}} } \right) - E\left( {\frac{{{k_0}}}{{{k_{Ft}}}}\sqrt {\frac{{i{{ \Delta }_1} + \mu }}{{ \Delta _1^2 + {\mu ^2}}}} ,\sqrt {\frac{{i\mu  + {{ \Delta }_1}}}{{i\mu  - {{\Delta }_1}}}} } \right)} \right),
\end{flalign}
\begin{flalign}
\label{eqS16}
\int\limits_0^{{k_0}/{k_{Ft}}} {\frac{{{x^2}}}{{\sqrt {{{\left( {{x^2} - \mu  + {E_g}} \right)}^2} +  \Delta _2^2} }}} dx = \sqrt {i{{\ \Delta }_2} + \mu  - {E_g}}  \times  \nonumber \\
\left( {F\left( {\frac{{{k_0}}}{{{k_{Ft}}}}\sqrt {\frac{{i{{\Delta }_2} + \mu  - {E_g}}}{{ \Delta _2^2 + {{\left( {\mu  - {E_g}} \right)}^2}}}} ,\sqrt {\frac{{i\left( {\mu  - {E_g}} \right) + {{\Delta }_2}}}{{i\left( {\mu  - {E_g}} \right) - {{ \Delta }_2}}}} } \right) - E\left( {\frac{{{k_0}}}{{{k_{Ft}}}}\sqrt {\frac{{i{{ \Delta }_2} + \mu  - {E_g}}}{{ \Delta _2^2 + {{\left( {\mu  - {E_g}} \right)}^2}}}} ,\sqrt {\frac{{i\left( {\mu  - {E_g}} \right) + {{ \Delta }_2}}}{{i\left( {\mu  - {E_g}} \right) - {{ \Delta }_2}}}} } \right)} \right),
\end{flalign}
\end{widetext}
where $F(z,\nu)$ and $E(z,\nu)$ are incomplete elliptic integrals of the first and the second kind and $i$ is the imaginary unit. This yields the system of equations for the energy gaps 
\begin{widetext}
\begin{flalign}
\label{eqS17}
{{ \Delta }_1}\sqrt {i{{\Delta }_1} + \mu } \left({F\left( {\frac{{{k_0}}}{{{k_{Ft}}}}\sqrt {\frac{{i{{ \Delta }_1} + \mu }}{{\tilde \Delta _1^2 + {\mu ^2}}}} ,\sqrt {\frac{{i\mu  + {{\Delta }_1}}}{{i\mu  - {{ \Delta }_1}}}}} \right) - E\left({\frac{{{k_0}}}{{{k_{Ft}}}}\sqrt {\frac{{i{{\Delta }_1} + \mu }}{{\Delta _1^2 + {\mu ^2}}}} ,\sqrt {\frac{{i\mu  + {{\tilde \Delta }_1}}}{{i\mu  - {{\Delta }_1}}}} } \right)} \right) = & \nonumber \\
\frac{{\frac{1}{{\frac{{{k_0}}}{{{k_{Ft}}}} - \frac{\pi }{{2{k_{F2}}{a_{22}}}}\frac{{{k_{F2}}}}{{{k_{Ft}}}}}}{{\Delta }_1} -  \frac{3}{{4}}{{\tilde U}_{12}}{{\left( {\frac{{{k_{Ft}}}}{{{k_0}}}} \right)}^2}{{ \Delta }_2}}}{{\frac{1}{{\frac{{{k_0}}}{{{k_{Ft}}}} + \frac{{{k_0}}}{{{k_{Ft}}}}\left( {\rho  - 1} \right) - \frac{\pi }{{2{k_{F2}}{a_{22}}}}\frac{{{k_{F2}}}}{{{k_{Ft}}}}\rho }} \cdot \frac{1}{{\frac{{{k_0}}}{{{k_{Ft}}}} - \frac{\pi }{{2{k_{F2}}{a_{22}}}}\frac{{{k_{F2}}}}{{{k_{Ft}}}}}} -  \frac{9}{{16}}{{\tilde U}_{12}}{{\tilde U}_{21}}{{\left( {\frac{{{k_{Ft}}}}{{{k_0}}}} \right)}^4}}},
\end{flalign}
\begin{flalign}
\label{eqS18}
{\Delta _2}\sqrt {i{\Delta _2} + \mu  - {E_g}} \left( {F\left( {\frac{{{k_0}}}{{{k_{Ft}}}}\sqrt {\frac{{i{\Delta _2} + \mu  - {E_g}}}{{\Delta _2^2 + {{\left( {\mu  - {E_g}} \right)}^2}}}} ,\sqrt {\frac{{i\left( {\mu  - {E_g}} \right) + {\Delta _2}}}{{i\left( {\mu  - {E_g}} \right) - {\Delta _2}}}} } \right) -  E\left( {\frac{{{k_0}}}{{{k_{Ft}}}}\sqrt {\frac{{i{\Delta _2} + \mu  - {E_g}}}{{\tilde \Delta _2^2 + {{\left( {\mu  - {E_g}} \right)}^2}}}} ,\sqrt {\frac{{i\left( {\mu  - {E_g}} \right) + {\Delta _2}}}{{i\left( {\mu  - {E_g}} \right) - {\Delta _2}}}} } \right)} \right) = & \nonumber \\
\frac{{\frac{1}{{\frac{{{k_0}}}{{{k_{Ft}}}} + \frac{{{k_0}}}{{{k_{Ft}}}}\left( {\rho  - 1} \right) - \frac{\pi }{{2{k_{F2}}{a_{22}}}}\frac{{{k_{F2}}}}{{{k_{Ft}}}}\rho }}{\Delta _2} - \frac{3}{4}{{\tilde U}_{21}}{{\left( {\frac{{{k_{Ft}}}}{{{k_0}}}} \right)}^2}{\Delta _1}}}{{\frac{1}{{\frac{{{k_0}}}{{{k_{Ft}}}} + \frac{{{k_0}}}{{{k_{Ft}}}}\left( {\rho  - 1} \right) - \frac{\pi }{{2{k_{F2}}{a_{22}}}}\frac{{{k_{F2}}}}{{{k_{Ft}}}}\rho }} \cdot \frac{1}{{\frac{{{k_0}}}{{{k_{Ft}}}} - \frac{\pi }{{2{k_{F2}}{a_{22}}}}\frac{{{k_{F2}}}}{{{k_{Ft}}}}}} - \frac{9}{{16}}{{\tilde U}_{12}}{{\tilde U}_{21}}{{\left( {\frac{{{k_{Ft}}}}{{{k_0}}}} \right)}^4}}}.
\end{flalign}
\end{widetext}
For large values of  $\frac{{{k_0}}}{{{k_{Ft}}}} \gg 1$  we expand in series incomplete elliptic integrals of the first and the second kind via full elliptic integrals of the 1st and 2nd kind  $F(\nu)$ and $E(\nu)$  
\begin{widetext}
\begin{eqnarray}
\label{eqS19}
F\left( {\frac{{{k_0}}}{{{k_{Ft}}}}\sqrt {\frac{{i{{\Delta }_1} + \mu }}{{\tilde \Delta _1^2 + {\mu ^2}}}} ,\sqrt {\frac{{i\mu  + {{\Delta }_1}}}{{i\mu  - {{\Delta }_1}}}} } \right) - E\left( {\frac{{{k_0}}}{{{k_{Ft}}}}\sqrt {\frac{{i{{\Delta }_1} + \mu }}{{\Delta _1^2 + {\mu ^2}}}} ,\sqrt {\frac{{i\mu  + {{\Delta }_1}}}{{i\mu  - {{ \Delta }_1}}}} } \right) \approx \frac{1}{{\sqrt {i{{\Delta }_1} + \mu } }}\frac{{{k_0}}}{{{k_{Ft}}}} + iK\left( {i\sqrt {\frac{{2{{\Delta }_1}}}{{i\mu  - {{\Delta }_1}}}} } \right) -  \nonumber \\
\sqrt {\frac{{i\mu  - {{ \Delta }_1}}}{{i\mu  + {{\Delta }_1}}}} \left( {\frac{{i\mu  + {{ \Delta }_1}}}{{i\mu  - {{\tilde \Delta }_1}}}E\left( {\sqrt {\frac{{i\mu  - {{ \Delta }_1}}}{{i\mu  + {{ \Delta }_1}}}} } \right) - \sqrt {\frac{{i\mu  + {{ \Delta }_1}}}{{i\mu  - {{\Delta }_1}}}} E\left( {\sqrt {\frac{{i\mu  + {{ \Delta }_1}}}{{i\mu  - {{\Delta }_1}}}} } \right) - \frac{{2{{ \Delta }_1}}}{{i\mu  - {{\Delta }_1}}}K\left( {\sqrt {\frac{{i\mu  - {{ \Delta }_1}}}{{i\mu  + {{\Delta }_1}}}} } \right)} \right).
\end{eqnarray}
\end{widetext}
The same asymptotic expansion can be performed for the left part of Eq. (\ref{eqS18}). For the right parts of Eqs. (\ref{eqS17}) and (\ref{eqS18}) expansion gives
\begin{widetext}
\begin{eqnarray}
\label{eqS20}
\frac{{\frac{1}{{\frac{{{k_0}}}{{{k_{Ft}}}} - \frac{\pi }{{2{k_{F2}}{a_{22}}}}\frac{{{k_{F2}}}}{{{k_{Ft}}}}}}{{ \Delta }_1} - \frac{3}{{4}}{{\tilde U}_{12}}{{\left( {\frac{{{k_{Ft}}}}{{{k_0}}}} \right)}^2}{{ \Delta }_2}}}{{\frac{1}{{\frac{{{k_0}}}{{{k_{Ft}}}} + \frac{{{k_0}}}{{{k_{Ft}}}}\left( {\rho  - 1} \right) - \frac{\pi }{{2{k_{F2}}{a_{22}}}}\frac{{{k_{F2}}}}{{{k_{Ft}}}}\rho }} \cdot \frac{1}{{\frac{{{k_0}}}{{{k_{Ft}}}} - \frac{\pi }{{2{k_{F2}}{a_{22}}}}\frac{{{k_{F2}}}}{{{k_{Ft}}}}}} - \frac{9}{{16}}{{\tilde U}_{12}}{{\tilde U}_{21}}{{\left( {\frac{{{k_{Ft}}}}{{{k_0}}}} \right)}^4}}} \approx \rho \frac{{{k_0}}}{{{k_{Ft}}}}{\Delta _1}-\frac{\pi }{{2{k_{F2}}{a_{22}}}}\frac{{{k_{F2}}}}{{{k_{Ft}}}}{\Delta_1}-\frac{3}{4}\rho {{\tilde U}_{12}}{\Delta _2},
\end{eqnarray}
\begin{eqnarray}
\label{eqS21}
\frac{{\frac{1}{{\frac{{{k_0}}}{{{k_{Ft}}}} + \frac{{{k_0}}}{{{k_{Ft}}}}\left( {\rho  - 1} \right) - \frac{\pi }{{2{k_{F2}}{a_{22}}}}\frac{{{k_{F2}}}}{{{k_{Ft}}}}\rho }}{{ \Delta }_2} - \frac{3}{{4}}{{\tilde U}_{21}}{{\left( {\frac{{{k_{Ft}}}}{{{k_0}}}} \right)}^2}{{\Delta }_1}}}{{\frac{1}{{\frac{{{k_0}}}{{{k_{Ft}}}} + \frac{{{k_0}}}{{{k_{Ft}}}}\left( {\rho  - 1} \right) - \frac{\pi }{{2{k_{F2}}{a_{22}}}}\frac{{{k_{F2}}}}{{{k_{Ft}}}}\rho }} \cdot \frac{1}{{\frac{{{k_0}}}{{{k_{Ft}}}} - \frac{\pi }{{2{k_{F2}}{a_{22}}}}\frac{{{k_{F2}}}}{{{k_{Ft}}}}}} - \frac{9}{{16}}{{\tilde U}_{12}}{{\tilde U}_{21}}{{\left( {\frac{{{k_{Ft}}}}{{{k_0}}}} \right)}^4}}} \approx  \frac{{{k_0}}}{{{k_{Ft}}}}{\Delta _2}-\frac{\pi }{{2{k_{F2}}{a_{22}}}}\frac{{{k_{F2}}}}{{{k_{Ft}}}}{\Delta_2}-\frac{3}{4}\rho {{\tilde U}_{21}}{\Delta _1},
\end{eqnarray}
\end{widetext}
After substitutions we finally obtain
\begin{widetext}
\begin{eqnarray}
\label{eqS22}
\frac{{i{\Delta }_1K\left( {i\sqrt {\frac{{2{{\Delta }_1}}}{{i\mu  - {{\Delta }_1}}}} } \right)}}{{\sqrt {i{{\tilde \Delta }_1} + \mu } }} - {\Delta }_1 \sqrt {i{{ \Delta }_1} + \mu } \sqrt {\frac{{i\mu  - {{ \Delta }_1}}}{{i\mu  + {{ \Delta }_1}}}} \left( {\frac{{i\mu  + {{ \Delta }_1}}}{{i\mu  - {{\Delta }_1}}}E\left( {\sqrt {\frac{{i\mu  - {{ \Delta }_1}}}{{i\mu  + {{ \Delta }_1}}}} } \right) - } \right. \nonumber \\
\left. { - \sqrt {\frac{{i\mu  + {{ \Delta }_1}}}{{i\mu  - {{ \Delta }_1}}}} E\left( {\sqrt {\frac{{i\mu  + {{ \Delta }_1}}}{{i\mu  - {{\tilde \Delta }_1}}}} } \right) - \frac{{2{{\Delta }_1}}}{{i\mu  - {{ \Delta }_1}}}K\left( {\sqrt {\frac{{i\mu  - {{ \Delta }_1}}}{{i\mu  + {{ \Delta }_1}}}} } \right)} \right) = \left( {\rho  - 1} \right) \frac{{{k_0}}}{{{k_{Ft}}}}{\Delta _1}-\frac{\pi }{{2{k_{F2}}{a_{22}}}}\frac{{{k_{F2}}}}{{{k_{Ft}}}}{\Delta_1}-\frac{3}{4}\rho {{\tilde U}_{12}}{\Delta _2},
\end{eqnarray}
\begin{eqnarray}
\label{eqS23}
\frac{{i{\Delta }_2K\left( {i\sqrt {\frac{{2{{\Delta }_2}}}{{i\left( {\mu  - {E_g}} \right) - {{ \Delta }_2}}}} } \right)}}{{\sqrt {i{{ \Delta }_1} + \mu  - {E_g}} }} - {\Delta }_2\sqrt {i{{\Delta }_2} + \mu  - {E_g}} \sqrt {\frac{{i\left( {\mu  - {E_g}} \right) - {{\Delta }_2}}}{{i\left( {\mu  - {E_g}} \right) + {{\Delta }_2}}}} \left( {\frac{{i\left( {\mu  - {E_g}} \right) + {{ \Delta }_2}}}{{i\left( {\mu  - {E_g}} \right) - {{\Delta }_2}}}E\left( {\sqrt {\frac{{i\left( {\mu  - {E_g}} \right) - {{\Delta }_2}}}{{i\left( {\mu  - {E_g}} \right) + {{ \Delta }_2}}}} } \right)} \right.  -  \nonumber \\
\left. { \sqrt {\frac{{i\left( {\mu  - {E_g}} \right) + {{ \Delta }_2}}}{{i\left( {\mu  - {E_g}} \right) - {{ \Delta }_2}}}} E\left( {\sqrt {\frac{{i\left( {\mu  - {E_g}} \right) + {{ \Delta }_2}}}{{i\left( {\mu  - {E_g}} \right) - {{ \Delta }_2}}}} } \right) - \frac{{2{{ \Delta }_2}}}{{i\left( {\mu  - {E_g}} \right) - {{ \Delta }_2}}}K\left( {\sqrt {\frac{{i\left( {\mu  - {E_g}} \right) - {{ \Delta }_2}}}{{i\left( {\mu  - {E_g}} \right) + {{\Delta }_2}}}} } \right)} \right) &  = \nonumber \\  -\frac{\pi }{{2{k_{F2}}{a_{22}}}}\frac{{{k_{F2}}}}{{{k_{Ft}}}}{\Delta_2}-\frac{3}{4}\rho {{\tilde U}_{21}}{\Delta _1},
\end{eqnarray}
\end{widetext}
One can see that for Eq. (\ref{eqS23}) of the system there is no dependence on the cut-off momentum value ${k_0}$  but for Eq. (\ref{eqS22}) this dependence is present excepting the case of coinciding intraband coupling strengths when $\rho  \ne 1$.

If we will follow the strategy for the description of BCS-BEC properties with the fixed coupling strength in the first band then after similar analytical calculations equations for the energy gaps at zero temperature transform to the form
\begin{widetext}
\begin{eqnarray}
\label{eqS24}
\frac{{i{\Delta }_1K\left( {i\sqrt {\frac{{2{{\Delta }_1}}}{{i\mu  - {{\Delta }_1}}}} } \right)}}{{\sqrt {i{{ \Delta }_1} + \mu } }} - {\Delta }_1 \sqrt {i{{ \Delta }_1} + \mu } \sqrt {\frac{{i\mu  - {{ \Delta }_1}}}{{i\mu  + {{\Delta }_1}}}} \left( {\frac{{i\mu  + {{\Delta }_1}}}{{i\mu  - {{\Delta }_1}}}E\left( {\sqrt {\frac{{i\mu  - {{\Delta }_1}}}{{i\mu  + {{ \Delta }_1}}}} } \right) - } \right.  \nonumber \\
\left. { - \sqrt {\frac{{i\mu  + {{\Delta }_1}}}{{i\mu  - {{\Delta }_1}}}} E\left( {\sqrt {\frac{{i\mu  + {{\Delta }_1}}}{{i\mu  - {{ \Delta }_1}}}} } \right) - \frac{{2{{\Delta }_1}}}{{i\mu  - {{ \Delta }_1}}}K\left( {\sqrt {\frac{{i\mu  - {{\Delta }_1}}}{{i\mu  + {{ \Delta }_1}}}} } \right)} \right) =-\frac{\pi }{{2{k_{F1}}{a_{11}}}}\frac{{{k_{F1}}}}{{{k_{Ft}}}}{\Delta_1}-\frac{3}{4} {{\tilde U}_{12}}{\Delta _2},
\end{eqnarray}
\begin{eqnarray}
\label{eqS25}
\frac{{i{\Delta }_2K\left( {i\sqrt {\frac{{2{{\Delta }_2}}}{{i\left( {\mu  - {E_g}} \right) - {{\Delta }_2}}}} } \right)}}{{\sqrt {i{{\Delta }_1} + \mu  - {E_g}} }} - {\Delta }_2\sqrt {i{{\Delta }_2} + \mu  - {E_g}} \sqrt {\frac{{i\left( {\mu  - {E_g}} \right) - {{\Delta }_2}}}{{i\left( {\mu  - {E_g}} \right) + {{\Delta }_2}}}} \left( {\frac{{i\left( {\mu  - {E_g}} \right) + {{\Delta }_2}}}{{i\left( {\mu  - {E_g}} \right) - {{\Delta }_2}}}E\left( {\sqrt {\frac{{i\left( {\mu  - {E_g}} \right) - {{\Delta }_2}}}{{i\left( {\mu  - {E_g}} \right) + {{\Delta }_2}}}} } \right)} \right.  \nonumber \\
\left. { - \sqrt {\frac{{i\left( {\mu  - {E_g}} \right) + {{ \Delta }_2}}}{{i\left( {\mu  - {E_g}} \right) - {{\Delta }_2}}}} E\left( {\sqrt {\frac{{i\left( {\mu  - {E_g}} \right) + {{\tilde \Delta }_2}}}{{i\left( {\mu  - {E_g}} \right) - {{\tilde \Delta }_2}}}} } \right) - \frac{{2{{ \Delta }_2}}}{{i\left( {\mu  - {E_g}} \right) - {{ \Delta }_2}}}K\left( {\sqrt {\frac{{i\left( {\mu  - {E_g}} \right) - {{\Delta }_2}}}{{i\left( {\mu  - {E_g}} \right) + {{ \Delta }_2}}}} } \right)} \right) &  = \nonumber \\ -\frac{\pi }{{2{k_{F2}}{a_{22}}}}\frac{{{k_{F2}}}}{{{k_{Ft}}}}{\Delta_2}-\frac{3}{4}{{\tilde U}_{21}}{\Delta _1},
\end{eqnarray}
\end{widetext}
where there is no cut-off momentum dependence and in comparison with  Eqs. (\ref{eqS22}) and (\ref{eqS23}) within this strategy the solution of Eqs. (\ref{eqS24}) and (\ref{eqS25}) is insensitive to the selection of $k_0$.

\end{document}